\documentclass[journal,10pt]{IEEEtran}
\usepackage{amssymb}
\usepackage{amsmath}
\usepackage{cite}
\usepackage{url}
\usepackage{xcolor}
\usepackage{cite,graphicx,amsmath,amssymb}
\usepackage{subfigure}
\usepackage{citesort}
\usepackage{fancyhdr}
\usepackage{mdwmath}
\usepackage{mdwtab}
\usepackage{caption}
\usepackage{amsthm}
\usepackage{algorithm}
\usepackage{algorithmic}
\usepackage{graphicx} 
\usepackage{epstopdf}
\usepackage{setspace}
\usepackage{tikz}
\usepackage{amsmath, bm}
\usepackage{caption}
\usepackage[justification=centering]{caption}
\usepackage{geometry}
\newtheorem{remark}{Remark}
\newtheorem{theorem}{Theorem}

\newtheorem{lemma}{Lemma}

\newtheorem{corollary}{Corollary}

\makeatletter
\def\ScaleIfNeeded{%
\ifdim\Gin@nat@width>\linewidth \linewidth \else \Gin@nat@width
\fi } \makeatother

\begin{document}

\title{\huge{Hybrid Reinforcement Learning for STAR-RISs: A Coupled Phase-Shift Model Based Beamformer}
}

\author{\normalsize {Ruikang~Zhong,~\IEEEmembership{\normalsize Graduate Student Member,~IEEE,}
Yuanwei~Liu,~\IEEEmembership{\normalsize Senior Member,~IEEE,}\\
Xidong~Mu,~\IEEEmembership{\normalsize Graduate Student Member,~IEEE,}
Yue~Chen,~\IEEEmembership{\normalsize Senior Member,~IEEE,}\\
Xianbin~Wang,~\IEEEmembership{\normalsize Fellow,~IEEE,}
Lajos~Hanzo,~\IEEEmembership{\normalsize Life Fellow,~IEEE.}
}

\thanks{Ruikang~Zhong, Yuanwei~Liu, and Yue~Chen are with the school of School of Electronic Engineering and Computer Science, Queen Mary University of London, London E1 4NS, U.K. (e-mail: r.zhong@qmul.ac.uk; yuanwei.liu@qmul.ac.uk; yue.chen@qmul.ac.uk).

Xidong~Mu is with  the School of Artificial Intelligence, Beijing University of Posts and Telecommunications, Beijing, 100876, China. He is also with the school of School of Electronic Engineering and Computer Science, Queen Mary University of London, London E1 4NS, U.K. (email:muxidong@bupt.edu.cn).

Xianbin Wang is with Department of Electrical and Computer Engineering, Western University, London, ON N6A5B9, Canada (e-mail: xianbin.wang@uwo.ca).

Lajos~Hanzo is with the School of Electronics and Computer Science, University of Southampton, Southampton SO17 1BJ, U.K. (e-mail:lh@ecs.soton.ac.uk).

R. Zhong would like to acknowledge the financial support of the China Scholarship Council (No.201908610187). L. Hanzo would like to acknowledge the financial support of the Engineering and Physical Sciences Research Council projects EP/P034284/1 and EP/P003990/1 (COALESCE) as well as of the European Research Council's Advanced Fellow Grant QuantCom (Grant No. 789028)

}
}

\maketitle
\vspace{-1.3cm}
\begin{abstract}

A simultaneous transmitting and reflecting reconfigurable intelligent surface (STAR-RIS) assisted multi-user downlink multiple-input single-output (MISO) communication system is investigated. In contrast to the existing ideal STAR-RIS model assuming an independent transmission and reflection phase-shift control, a practical coupled phase-shift model is considered. Then, a joint active and passive beamforming optimization problem is formulated for minimizing the long-term transmission power consumption, subject to the coupled phase-shift constraint and the minimum data rate constraint. Despite the coupled nature of the phase-shift model, the formulated problem is solved by invoking a hybrid continuous and discrete phase-shift control policy.
Inspired by this observation, a pair of hybrid reinforcement learning (RL) algorithms, namely the hybrid deep deterministic policy gradient (hybrid DDPG) algorithm and the joint DDPG \& deep-Q network (DDPG-DQN) based algorithm are proposed. The hybrid DDPG algorithm controls the associated high-dimensional continuous and discrete actions by relying on the hybrid action mapping. By contrast, the joint DDPG-DQN algorithm constructs two Markov decision processes (MDPs) relying on an inner and an outer environment, thereby amalgamating the two agents to accomplish a joint hybrid control.
Simulation results demonstrate that the STAR-RIS has superiority over other conventional RISs in terms of its energy consumption. Furthermore, both the proposed algorithms outperform the baseline DDPG algorithm, and the joint DDPG-DQN algorithm achieves a superior performance, albeit at an increased computational complexity.

\end{abstract}

\section{Introduction}

Each generation of wireless networks has aimed for improving the quality of service (QoS)~\cite{8922634}. While the fifth-generation (5G) network is being rolled out, investigators turned to the next-generation (6G) research both in academia and industry. In this context, reconfigurable intelligent surfaces (RISs) have emerged as a competitive 6G component \cite{9139273,9086766}. They constitute a low-cost solution for improving the propagation conditions for edge users \cite{9521988} and millimeter wave (mmWave) communications \cite{9520407}, as well as energy-efficient communication \cite{8741198}. Furthermore, RISs can also provide additional functions beyond signal enhancement, such as integrated sensing and reflecting \cite{alamzadeh2021reconfigurable}, channel estimation \cite{9593172}, and user localization \cite{9528041}.

Due to their similar functions and roles, the performance of RISs and wireless relays has been comprehensively compared in \cite{9530717,9122596}. Despite the appealingly low complexity and noise figure \cite{9122596}, an undeniable fact is that the reflecting-only RISs are likely to have a coverage disadvantage compared to omnidirectional relays due to their $180^\circ$ half-plane reflection limitation. However, in practical scenarios, users roam at both sides of the RIS, and the reflecting-only RIS cannot provide signal enhancements for users located at the back of the RIS. This deficiency of reflecting-only RISs inspires the design of double-sided RISs to achieve $360^\circ$ coverage boundary.

Hence, a simultaneous transmitting and reflecting reconfigurable intelligent surface (STAR-RIS) model emerged as an ameliorated version of the reflecting-only RIS \cite{mu2021simultaneously}. As an advanced derivative, an additional transmission function is facilitated by the simultaneous transmitting and reflecting (STAR) elements. The STAR-RIS accommodate a number of STAR elements on a surface to separate the incident electromagnetic waves into transmitted and reflected signals. Consequently, the transmitted signals and reflected signals can form a pair of sectors on both sides of the surface simultaneously, thereby solving the limitations of RISs in terms of their coverage region. In contrast to the reflecting-only RIS, which has to be in the vicinity of either the base station (BS) or the users, STAR-RIS may be positioned flexibly. Furthermore, if the BS or users are out of the optimal orientation of the RIS, the gain provided by the reflecting-only RIS deteriorates distinctly \cite{9201413}. Fortunately, the emergence of STAR-RISs can relax the orientation requirement of (STAR) RISs\footnote{The abbreviation '(STAR) RISs' refers to both reflecting-only RISs and STAR-RISs.}, since the transmission and reflection sectors have the opposite orientation.

\subsection{The State of the Art}
Given its short history, the research of STAR-RIS is still in its infancy. Zhu \textit{et al.} \cite{zhu2014dynamic,7274678} presented the design of metasurfaces to control the amplitudes and phase-shifts of the transmission (refraction) and reflection of electromagnetic waves. The implementation of STAR elements was discussed in \cite{zhu2014dynamic}, which laid a foundation of STAR-RISs. While Xu \textit{et al.} \cite{9437234} proposed a physical model for the STAR-RIS and demonstrated that it is feasible to extend reflecting-only RISs to STAR-RISs. Three protocols were conceived for STAR-RISs in \cite{mu2021simultaneously}, namely the energy splitting, mode switching, and time switching models. The phase-shift optimization problem of STAR-RISs was investigated in \cite{9200683}, where a branch-and-bound based algorithm was proposed for maximizing the downlink's spectral efficiency. The coverage performance of STAR-RIS networks was investigated by Wu \textit{et al.} \cite{9462949}, and a one-dimensional search-based algorithm was proposed for optimizing the coverage of STAR-RISs. The simulation results in \cite{9462949} confirmed that STAR-RISs exhibit significant coverage advantage over reflecting-only RISs. By considering a coupled transmitting and reflecting model, Liu \textit{et al.} \cite{liu2021simultaneously} proposed an efficient element-wise alternating optimization algorithm for minimizing the transmit power of a STAR-RIS assisted non-orthogonal multiple access (NOMA) network. Unfortunately, although several STAR-RIS models have been proposed in the aforementioned research, these models assume arbitrary transmission coefficients (TCs) and reflection coefficients (RCs) for the STAR elements. However, as pointed out in \cite{9115725}, a passive device such as the STAR-element is not likely to provide arbitrary phase and amplitude responses.


Deep learning (DL) and deep reinforcement learning (DRL) are capable of intelligently managing RISs \cite{wang2020thirty}. DL is primarily used for channel estimation in RIS-assisted networks, in order to extract the channel state information (CSI) in support of RIS-based passive beamforming. Gao \textit{et al.} \cite{9367208} employed a synthetic deep neural network (DNN) for sequentially estimating the BS-RIS channel and RIS-user channel, while reducing the pilot overhead and guaranteeing the estimation accuracy. DL was also invoked for the phase-shift control of RISs. For example, an unsupervised DL algorithm was proposed for joint active and passive beamforming optimization in \cite{9274528}. In terms of DRL, by adopting a deep deterministic policy gradient (DDPG) algorithm, Huang \textit{et al.} \cite{9110869} conceived the joint optimization of the transmit beamforming matrix of the BS and the phase-shift matrix of the RIS. Yang \textit{et al.} \cite{9206080} proposed a deep Q network (DQN) for secure beamforming guarding against eavesdroppers in dynamic environments. Their simulation results verified that the DQN approach is capable of enhancing the secrecy rate and the satisfaction probability of users. Furthermore, a DRL scheme was invoked in \cite{9174801} for optimizing a RIS-assisted NOMA network, where a long short-term memory (LSTM) based echo state network (ESN) algorithm collaborated with a decaying double deep Q-network (D3QN) for intelligently controlling the RIS according to the users' data demand. In \cite{9371415}, a proximal policy optimization (PPO) algorithm was developed for minimizing the expected Age-of-Information (AoI) for an aerial RIS. Specifically for the STAR-RIS scenario, a federated learning algorithm was proposed in \cite{ni2021star} for maximizing the achievable data rate of a STAR-RIS assisted heterogeneous NOMA network.

\subsection{Motivations}

Although STAR-RISs have the aforementioned advantages, the effective design of the RC and TC of STAR-RISs has become a new challenge. Firstly, the STAR-RIS requires joint transmission and reflection beamforming, which is exceedingly more complex than reflection-only beamforming. What aggravates the situation further is that the STAR-RISs cannot independently adjust the TCs and RCs in practice, since the electric and magnetic impedances are unlikely to leave arbitrary values, but they depend on the electromagnetic properties of the STAR elements \cite{zhu2014dynamic}. Furthermore, the coupling of the TCs and RCs requires a hybrid continuous and discrete control scheme for the phase-shift design. Given the above-mentioned adversities, it is a challenge to jointly solve the transmission and reflection beamforming problem for STAR-RISs, especially considering that the existing convex optimization and machine learning solutions basically only support either continuous or discrete control. Although several hybrid algorithms have been proposed in the field of computer science~\cite{neunert2020continuous,delalleau2019discrete,li2021hyar}, they are designed for minuscule action dimensions. For example, only four discrete actions were assumed in \cite{delalleau2019discrete} since gaming controllers usually have four buttons. However, the possible number of actions can be $a^N$ for the STAR-RIS scenario, where $a$ represents the possible number of actions for a single STAR element and $N$ represents the total number of elements employed. Based on the current assumptions, prototypes of (STAR) RISs are likely to have a massive number of elements, which implies that the action dimension of STAR-RISs substantially exceeds the design in existing algorithms. Since there are lacking suitable hybrid algorithms for the optimization of STAR-RISs, we propose two hybrid reinforcement learning (RL) algorithms for joint active and passive beamforming design for the BS and the STAR-RIS.

\subsection{Our contributions}

In this paper, we are committed to establishing a practical STAR-RIS model having coupled TCs and RCs, which can provide basis for potential research. According to the proposed model, we embark upon the investigation and optimization of the performance of STAR-RISs. Specifically, we provide the following contributions.

\begin{itemize}

\item We conceive a STAR-RIS model for broadening the coverage of reflecting-only RISs. Specifically, the practical electromagnetic property of STAR elements are considered, resulting in a coupled phase-shift of the transmission and reflection. Based on the proposed model, a joint active and passive beamforming problem that requires hybrid control for the phase-shift and amplitude is formulated for minimizing the long-term power consumption.

\item We propose a hybrid DDPG algorithm for solving the hybrid control problem caused by the energy splitting nature of STAR elements. The hybrid control is carried out by mapping each output node of the actor network to the transmission and reflection actions of each STAR element. The proposed hybrid DDPG solution provides high-dimensional continuous and discrete phase-shift optimization for STAR-RISs.

\item We develop a joint DDPG-DQN algorithm as a high performance-solution.  The joint DDPG-DQN scheme can handle hybrid control by employing two collaborated agents, where a DDPG agent is in charge of the continuous control and a DQN agent is invoked for the discrete control.

\item The performance of the STAR-RIS relying on the proposed algorithms is evaluated by computer simulation revealing that it outperforms both the reflecting-only and the double spliced RISs.  Furthermore, the hybrid DDPG algorithm outperforms its plain DDPG counterpart without increasing its complexity, while the joint DDPG-DQN algorithm attains optimality at an increased complexity.

\end{itemize}

\subsection{Organizations}

Section \ref{section:2} illustrates the model of our STAR-RIS assisted wireless network, including the coupled phase-shift model, channel model, signal model, and the problem formulation. Section \ref{section:3} introduces the hybrid DDPG algorithm as a low complexity solution, while Section \ref{section:4} presents the high-performance amalgamated DDPG-DQN algorithm. Section  \ref{section:5} provides the performance analysis of the proposed framework and algorithms. Section \ref{section:6} concludes the paper.

\begin{table}[t!]
 \caption{Key Notations}\label{SP}
 \footnotesize
 \renewcommand\arraystretch{1.2}
 \begin{tabular}{|l|r|}
  \hline
  Parameter & Description     \\
  \hline
  $M,N$ & number of antenna and STAR elements   \\
  \hline
  $K,U,I$ & number of total users, $\mathcal{R}$ users and $\mathcal{T}$ users   \\
  \hline
  $\beta_{\mathcal{R},n}$ & amplitude response for reflection   \\
  \hline
  $\mathbf{\Theta }_\mathcal{R},\mathbf{\Theta }_\mathcal{T}$ & phase shift of for reflection and transmission  \\
  \hline
  $\bm{H}_{b,\mathcal{R}} , \bm{H}_{b,\mathcal{T}} $  & channel from the BS to $\mathcal{R}\&\mathcal{T}$ users  \\
  \hline
  $\bm{H}_{b,r}$ & channel from the BS to the STAR-RIS  \\
  \hline
  $\bm{H}_{r,\mathcal{R}} , \bm{H}_{r,\mathcal{T}} $ & channel from the STAR-RIS to $\mathcal{R}\&\mathcal{T}$ users   \\
  \hline
  $\mathbf{w}_{k,t}$ & beamforming vector for user $k$ at time $t$   \\
  \hline
  $\mathbf{x}_{b,t}$ & transmitted signal at the BS   \\
  \hline
  $\mathbf{x}_{r,t}$ & received signal at the STAR-RIS   \\
  \hline
  $y_{u,t}, y_{i,t}$ & received signal at user $\mathcal{R}_u$ and $\mathcal{T}_i$   \\
  \hline
  $\gamma_{u,t}, \gamma_{i,t}$ & SINR at user $\mathcal{R}_u$ and $\mathcal{T}_i$   \\
  \hline
  $R_{k,t}$ & achievable data rate for user $k$   \\
  \hline
  $\mathbf{s}_t$ & state for DRL agents   \\
  \hline
  $\mathbf{a}^c_t,\mathbf{a}^d_t$ & continuous and discrete actions   \\
  \hline
  $\mathbf{a}^h_t$ & hybrid actions   \\
  \hline
  $r_t$ & reward for DRL agent   \\
  \hline
  $\bm{\omega}^\mu_t,\bm{\omega}^Q_t$ & parameters for actor and critic networks  \\
  \hline
  $\bm{\omega}^\mathcal{Q}_t$ & parameters for the deep Q network \\
  \hline
  $\bm{\omega}^{\mu'}_t,\bm{\omega}^{Q'}_t,\bm{\omega}^{\mathcal{Q}'}_t$ & parameters for target networks \\
  \hline
 \end{tabular}
\end{table}

\section{System Model}\label{section:2}


\subsection{Model of STAR-RISs}

We employ an energy splitting model for supporting simultaneous transmission and reflection~\cite{mu2021simultaneously}, where the STAR-RIS is capable of splitting the incident signal into the transmitted and reflected signals, partitioning the space into the transmission and reflection zones. Mobile users can be served by the transmitted or reflected signal, respectively, depending on which region they happen to be roaming in. In order to perform joint beamforming to covering both the transmission and reflection sectors, the TC and RC of each STAR element have to be appreciatively integrated, which are denoted as $\beta_{\mathcal{R},n} e^{j\theta_{\mathcal{R},n}}$ and $\beta_{\mathcal{T},n} e^{j\theta_{\mathcal{T},n}}, n = 1,2,...,N$.

It is worth noting that for any STAR element, the TCs an RCs are determined by its resistance and reactance. Therefore, it is non-trivial to independently adjust the coefficients.  For a given RC of $\beta_{\mathcal{R},n} e^{j\theta_{\mathcal{R},n}}$, according to the conservation of energy\footnote{We assume that there is no energy loss in the process of transmission and reflection. The energy loss has an impact on the phase shift relationship between transmission and reflection \cite{zhu2014dynamic}, which would make the beamforming design more complex. We would consider the non-conservation model in our future work.}, we have $\beta_{\mathcal{T},n} = \sqrt{1- \beta_{\mathcal{R},n}^2}$. Then, simplifying RC $\beta_{\mathcal{R},n}$ as $\beta_{n}$, the TC can be calculated as $\sqrt{1- \beta_{n}^2} e^{j\theta_{\mathcal{T},n}}$. As pointed out in \cite{zhu2014dynamic}, for STAR elements, the coupling between the TC's phase-shift $\theta_{\mathcal{T},n}$, the RC's phase-shift $\theta_{\mathcal{R},n}$ and the amplitude $\beta_n$ follows a relationship as
\begin{align}
\beta_{n} \sqrt{1- \beta_{n}^2} \text{ cos}(\theta_{\mathcal{R},n} - \theta_{\mathcal{T},n}) = 0.
\end{align}

Thus, for a STAR-RIS having $N$ elements, the transmission and reflection matrices have a diagonal structure given by
\begin{align}
{\mathbf{\Theta }_\mathcal{R}} = {\rm{diag}}\left( {{\beta_{1} e^{j{\theta_{\mathcal{R},1}}}},{\beta_{2} e^{j{\theta _{\mathcal{R},2}}}}, \cdots ,{\beta_{N} e^{j{\theta _{\mathcal{R},N}}}}} \right){\text{}},
\end{align}
\begin{align}
{\mathbf{\Theta }_\mathcal{T}} = {\rm{diag}}\left( {{\sqrt{1- \beta_{1}^2} e^{j{\theta_{\mathcal{T},1}}}}, \cdots ,{\sqrt{1- \beta_{N}^2} e^{j{\theta _{\mathcal{T},N}}}}} \right){\text{}}.
\end{align}



\subsection{System Description}\label{section:2A}

\begin{figure*}[t!] 
\centering 
\includegraphics[width=0.7\textwidth]{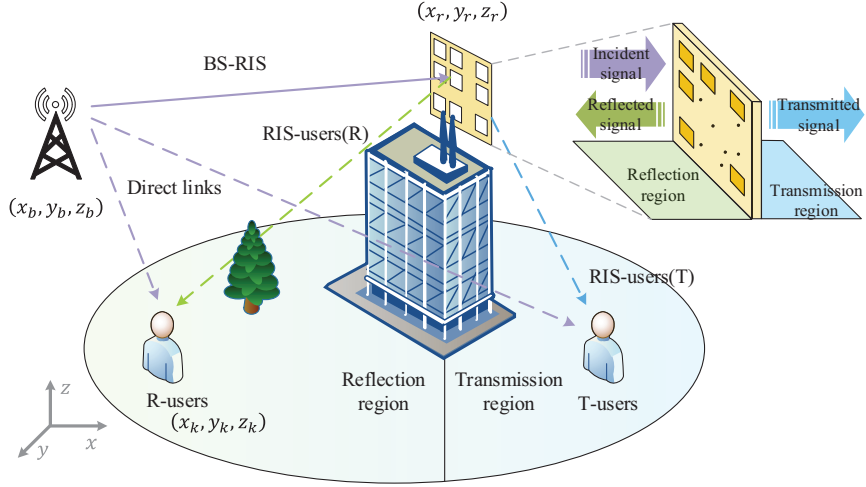} 
\caption{System model of STAR-RIS assisted wireless networks} 
\label{Fig.system} 
\end{figure*}

We consider a downlink scenario of Fig.~\ref{Fig.system}, where the AP is equipped with $M$ antennas, and the STAR-RIS has $N$ STAR elements. There are $K$ randomly roaming users and each having a single antenna. Each of the $N$ STAR elements has the amplitude response $\beta_n, n = 1,2,...,N$. We denote the locations of the BS, RIS, and users as ${\left( {{x_{b}},{y_{b}},{z_{b}}} \right)^T}$, ${\left( {{x_{r}},{y_{r}},{z_{r}}} \right)^T}$, and ${\left( {{x_{k}},{y_{k}},{z_{k}}} \right)^T}$, respectively. The STAR-RIS naturally partitions the users into two groups according to their locations. The users between the BS and the STAR-RIS receive direct signals from the BS and reflected signals from the STAR-RIS. This fraction of the users in the reflective region of the STAR-RIS are denoted by $\mathcal{R}_u$. For simplicity, we term this set of users as $\mathcal{R}$ users in the rest of the text. Correspondingly, the users served by direct BS signals and transmitted STAR-RIS signals is denoted by $\mathcal{T}_{i}$. The number of users obeys $K=U+I$, where $U$ and $I$ are the numbers of $\mathcal{R}$ users and $\mathcal{T}$ users\footnote{For simplicity of equations, the subscript $k = 1,2,...,K$ refers to any user, user $\mathcal{R}_u,u = 1,2,...,U$ or $\mathcal{T}_{i},i= 1,2,...,I$ refers to a $\mathcal{R}$ user or $\mathcal{T}$ user.}.

\subsection{Channel Model}\label{section:2C}

In the STAR-RIS scenario, multiple channels have to be considered, including the BS to STAR-RIS channel $\bm{H}_{b,r} \in {\mathbb{C}^{M \times N}}$, the direct channel spanning the BS to $\mathcal{R}$ and $\mathcal{T}$ users $\bm{H}_{b,\mathcal{R}} \in {\mathbb{C}^{M \times U}}, \bm{H}_{b,\mathcal{T}} \in {\mathbb{C}^{M \times I}}$, and the channel impinging from the STAR-RIS to $\mathcal{R}$ and $\mathcal{T}$ users $\bm{H}_{r,\mathcal{R}} \in {\mathbb{C}^{N \times U}}, \bm{H}_{r,\mathcal{T}} \in {\mathbb{C}^{N \times I}}$. For each specific user $\mathcal{R}_u$ and $\mathcal{T}_i$, the direct channels, and the $\mathcal{T}\&\mathcal{R}$ channels can be denoted as $\bm{h}_{b,u} \in {\mathbb{C}^{M \times 1}}, \bm{h}_{b,i} \in {\mathbb{C}^{M \times 1}}, \bm{h}_{r,u} \in {\mathbb{C}^{N \times 1}}$ and $\bm{h}_{r,i} \in {\mathbb{C}^{N \times 1}}$, respectively.

All channels are assumed to follow the quasi-static block fading model, where the fading coefficient remains constant in each time slot (TS) $t$. We assume that the channel $\bm{H}_{b,r}$ has a line-of-sight (LoS) path and obeys to the Rician distribution, since the BS and STAR-RIS have a LoS component owing to their selected positions. Thus, upon considering the path loss and the small scale fading, the Rician channel can be formulated as
\begin{align}
{\bm{H}_{b,r,t}}\!=\!\!\sqrt {\mathcal{L}_{b,r,t}\left( {{d_{b,r,t},f_c}} \right)}\! \left(\!\!{\sqrt {\frac{{{K}}}{{{K} + 1}}} {{\bm{L}_{b,r,t}}}\!\!+\!\!\sqrt {\frac{1}{{{K} + 1}}} {{\bm{G}_{b,r,t}}}  }\!\!\right),
\end{align}
where $\mathcal{L}_{b,r,t}$ represents the pathloss in the power domain, $K$ is the Rician factor, $\mathbf{G}_{b,r,t}\sim \mathcal{CN}(0,1)$ denotes the scattered paths, and ${{\bm{L}_{b,r,t}}}$ represents the LoS path between the BS and the RIS. Furthermore, we have ${{\bm{L}_{b,r,t}}} = \bm{a}_r(\varphi^{A},\psi^A) \bm{a}_{b}(\varphi^{D})^H , \bm{a}_{b}\in \mathbb{C}^{M\times1}, \bm{a}_r \in \mathbb{C}^{N\times1}$, where $\varphi^{A/D} \in [0,2\pi)$ and $\psi^{A} \in [-2/\pi,2/\pi)$ represent the azimuth and elevation angle-of-arrival (AoA)/angle-of-departure (AoD), $ \varphi^A = \arcsin(\frac{{y_r} - {y_{b}}}{\sqrt {{{\left( {{x_r} - {x_{b}}} \right)}^2} + {{\left( {{y_r} - {y_{b}}} \right)}^2}} })$, $\varphi^D = \pi/2 - \varphi^A$, and $\psi^A = \arcsin(\frac{{{z_r} - {z_b}}}{{\sqrt {{{\left( {{x_r} - {x_b}} \right)}^2} + {{\left( {{y_r} - {y_b}} \right)}^2}} }})$.

We assume that the BS has linear array antennas and the STAR RIS has a uniform planar array of reflective elements, where the antenna space and the reflective element space are $d_a$ and $d_e$. According to \cite{9110912}, we have
\begin{align}
a_{b}[m] =& e^{j2\pi(m-1)d_a\sin(\varphi^D)/\lambda},\\
a_r[n]=& e^{j2\pi(n-1)d_e(\lfloor n/N_x\rfloor\sin(\varphi^A)\sin(\psi^A)} \nonumber\\ &e^{(n-\lfloor n/N_x\rfloor N_x)\sin(\varphi^A)\cos(\psi^A))/\lambda},
\end{align}
where $M_x$ denotes the number of reflective elements in each row and $\lfloor\rfloor$ is the floor function.

The pathloss $\mathcal{L}$ follows the urban propagation model presented in 3GPP specification TR 36.873 \cite{TR36873}. For the path with LoS, the path loss can be given by
\begin{align}
\mathcal{L}_\text{LoS}\left( d,f_c \right) = 22.0\log_{10}{d} + 28.0 + 20\log_{10}{f_c},
\end{align}
where $d$ represents the 3D distance between the transmitter and the receiver, while $f_c$ is the carrier frequency. For the NLoS propagation, the path loss is given by
\begin{align}
\mathcal{L}_\text{NLoS} = \max[\mathcal{L}_\text{LoS}\left( d,f_c \right),\mathcal{L}_\text{NLoS}\left( d,f_c \right)],
\end{align}
\begin{align}
\mathcal{L}_\text{NLoS}\left( d,f_c \right) = &36.7\log{10}{d}+22.7 \nonumber \\&+26\log{10}{f_c}-0.3(z_r-1.5).
\end{align}

On the other hand, due to the random movements of users, LoS propagation may not necessarily be guaranteed, regardless whether the transmitting side is the BS or the STAR-RIS. Therefore, $\bm{H}_{b,{\mathcal{R}},t}, \bm{H}_{b,\mathcal{T},t}, \bm{H}_{r,{\mathcal{R}},t}$ and $\bm{H}_{r,{\mathcal{T}},t}$ are assumed to be NLoS channels and follow Rayleigh fading, which can be expressed as $\bm{H}_{b,{\mathcal{R}},t} \!=\! \sqrt{\mathcal{L}_{b,\mathcal{R},t}}\bm{G}_{b,{\mathcal{R}},t}, \bm{H}_{b,{\mathcal{T}},t} \!=\! \sqrt{\mathcal{L}_{b,\mathcal{T},t}}\bm{G}_{b,{\mathcal{T}},t}, \bm{H}_{r,{\mathcal{R}},t} \!=\! \sqrt{\mathcal{L}_{r,\mathcal{R},t}}\bm{G}_{r,{\mathcal{R}},t}, \bm{H}_{r,{\mathcal{T}},t} \!=\! \sqrt{\mathcal{L}_{r,\mathcal{T},t}}\bm{G}_{r,{\mathcal{T}},t}$.


\subsection{Signal Model}\label{section:3D}

We denote the information sequence and the active beamforming vectors for user $k$ at the BS by $s_{k,t}$ and $\mathbf{w}_{k,t}, \in {\mathbb{C}^{M \times K}}$.
The signal transmitted at TS $t$ can be expressed as
\begin{align}\label{xt}
\mathbf{x}_{b,t} = \sum\limits_{k = 1}^{K}  \mathbf{w}_{k,t} s_{k,t}.
\end{align}

Then, the incident signal at the STAR-RIS is given by
\begin{align}
\mathbf{x}_{r,t} =  \bm{H}_{b,r,t} \sum\limits_{k = K}^{U}  \mathbf{w}_{k,t} s_{k} + n_0,
\end{align}
where $n_0$ represents the Gaussian noise, and the received signal of user $\mathcal{R}^u$ is given by
\begin{align}
y_{u,t} =  [\bm{h}_{b,u,t} + \bm{h}_{r,u,t} \mathbf{\Theta}_{\mathcal{R},t} \bm{H}_{b,r,t}] \sum\limits_{u = 1}^{U}  \mathbf{w}_{u,t} s_{u,t} + n_0.
\end{align}

Correspondingly, the received signal of user $\mathcal{T}^i$ can be represented in a similar form as
\begin{align}
y_{i,t} =  [\bm{h}_{b,i,t} + \bm{h}_{r,i,t} \mathbf{\Theta}_{\mathcal{T},t} \bm{H}_{b,r,t}] \sum\limits_{i = 1}^{I}  \mathbf{w}_{i,t} s_{i,t} + n_0.
\end{align}

Given the received signal, the  signal-to-interference-plus-noise ratio (SINR) of user $\mathcal{R}^u$ and $\mathcal{T}^i$ is given by
\begin{align}\label{SINROMA}
\gamma _{u,t} = \frac{\mid [\bm{h}_{b,u,t} + \bm{h}_{r,u,t} \mathbf{\Theta}_{\mathcal{R},t} \bm{H}_{b,r,t}] \mathbf{w}_{u,t} \mid^2} {\mid[\bm{h}_{b,u,t} + \bm{h}_{r,u,t} \mathbf{\Theta}_{\mathcal{R},t} \bm{H}_{b,r,t}] \sum\limits_{k \leq K, k\neq u}  \mathbf{w}_{k,t} \mid^2 + \sigma^2},
\end{align}
\begin{align}\label{SINROMA}
\gamma _{i,t} = \frac{\mid [\bm{h}_{b,i,t} + \bm{h}_{r,i,t} \mathbf{\Theta}_{\mathcal{T},t} \bm{H}_{b,r,t}] \mathbf{w}_{i,t} \mid^2} {\mid[\bm{h}_{b,i,t} + \bm{h}_{r,i,t} \mathbf{\Theta}_{\mathcal{T},t} \bm{H}_{b,r,t}] \sum\limits_{k \leq K, k\neq i}  \mathbf{w}_{k,t} \mid^2 + \sigma^2},
\end{align}
where $\sigma^2$ represents the noise power. Therefore, given a bandwidth $B$, the achievable data rate of each user is given by
\begin{align}
R_{k,t} = B\log_2\left( {1 + \gamma _{k,t}} \right).
\end{align}

\subsection{Problem Formulation}\label{section:3E}

We aim for minimizing the power consumption of the BS by jointly optimizing the beamforming vector $\mathbf{w}_k$ for the BS and the TCs as well as RCs of the STAR-RIS. Again, the TCs and RCs of the STAR-RIS can be represented by the phase-shifts $\bm{\Theta}_\mathcal{T}, \bm{\Theta}_\mathcal{R}$, and amplitude coefficients $\bm{\beta}$. Therefore, the optimization problem can be formulated as
\begin{subequations}
\begin{align}\label{OPP}
&\min_{\mathbf{w},\bm{\Theta}_\mathcal{T},\bm{\Theta}_\mathcal{R},\bm{\beta}} \sum_{t=1}^{T} \sum_{k=1}^{K} \parallel \mathbf{w}^2_{k,t}\parallel,\\
\textrm{s.t.} \ \
& -\pi \leq \theta_{\mathcal{T},n,t} \le \pi, \forall n, \forall t,\label{OPPB}\\
& -\pi \leq \theta_{\mathcal{R},n,t} \le \pi, \forall n, \forall t,\label{OPPC}\\
& R_{k,t} \geq R_\text{QoS}, \forall k, \forall t,\label{OPPD}\\
& 0 < \beta_{n,t}\leq 1, \forall n, \forall t,\label{OPPE}\\
& \beta_{n,t} \sqrt{1- \beta_{n,t}^2} \cos(\theta_{\mathcal{R},n,t} - \theta_{\mathcal{T},n,t}) = 0, \label{OPPF}\\
& P_{b,t} \leq P_{\text{max}}, \label{OPPG}
\end{align}
\end{subequations}
where constraint \eqref{OPPB} and \eqref{OPPC} represent the legitimate range of the TC and RC phase-shifts. Constraint \eqref{OPPD} is a QoS constraint specifically the minimum data rate. Since (STAR) RISs are passive devices, their amplitude response is limited by the conservation of energy, as shown in \eqref{OPPE}. Constraint \eqref{OPPF} characterizes the phase and amplitude relationship of the TCs and RCs. Finally, \eqref{OPPG} is the maximum power constraint for the BS.

The challenge of solving our problem is not only owing to the joint consideration of TCs and RCs, but also due to the constraint \eqref{OPPE}.
Given the coupling between $\theta_{\mathcal{T},n}$ and $\theta_{\mathcal{R},n}$, the STAR elements $n$ cannot have independent arbitrary TCs and RCs. Furthermore, although the STAR-RIS aspire continuous phase-shift control for both transmission and reflection, one party of them can only have dualistic options. For example, assuming $\beta_n \neq 0$\footnote{If $\beta_n = 0$ or $\beta_n =1$, the STAR-RIS operates in either the full transmission or full reflection mode, which is not a preferred mode.}, once the TC is determined as $\theta_{\mathcal{T},n}$, the RC can only select the phase-shift from $\{\theta_{\mathcal{T},n}+\frac{\pi}{2},\theta_{\mathcal{T},n}-\frac{\pi}{2} \}$. Therefore, the coupled phase-shift model of STAR-RISs requires hybrid continuous and discrete control for the transmission and reflection, which motivates us to develop hybrid DRL algorithms for solving this challenge.

\section{Solution I: The Hybrid DDPG Algorithm}\label{section:3}

The DDPG algorithm was shown to constitute an efficient solution for continuous control problems \cite{lillicrap2015continuous}. For applying DRL approaches to solve our (STAR) RIS optimization problem, the transmission period has to obey a Markov decision process (MDP) \cite{li2021joint}. In TS $t \in T$, by checking the CSI of the current channels, the agent determines the current state $\mathbf{s}_t \in \mathbf{S}$ and decides to carry out the action $\mathbf{a}_t \in \mathbf{A}$, where $\mathbf{S}$ and $\mathbf{A}$ represent the state space and action space. The action refers to a vector storing the active and passive beamforming coefficients at the BS and STAR-RIS.

Since passive and active beamforming are jointly considered, the MDP state for each TS $t$ includes the CSI of the BS to STAR-RIS, BS to users (both $\mathcal{R}$ and $\mathcal{T}$ users), and the STAR-RIS to users channels, as shown in Fig. \ref{DDPG}.  Thus, $\mathbf{s}_t$ is given by
\begin{align}\label{S}
\mathbf{s}_t =\{\bm{H}_{b,r,t},\bm{H}_{b,\mathcal{R},t},\bm{H}_{b,\mathcal{T},t},\bm{H}_{r,\mathcal{R},t},\bm{H}_{r,\mathcal{T},t} \}.
\end{align}

Once action $\mathbf{a}_t$ is executed, the agent has to determine the reward $r_t$ according to the data rate and power consumption of the transceiver, and then the state would be constrained to $\mathbf{s}_{t+1} \in \mathbf{S}$. Once the above steps are completed, $(\mathbf{s}_t, \mathbf{a}_t, r_t, \mathbf{s}_{t+1})$ would be saved as a Markov transition in the replay buffer for the agent's training.

\subsection{DDPG training}\label{section:3A}

The objective of the DRL based agent training is to find the specific action $\mathbf{a}_t$ for each state $\mathbf{s}_t$, which maximizes the expected accumulated reward $\mathbb{E}[\sum\nolimits_{i = t}^{T} \gamma r_{t+1}]$, where $\gamma$ represents the discount factor $\gamma \in [0,1]$.
For a DDPG agent, the Q value of action $\mathbf{a}_t$ can be quantified by the Bellman equation of \cite{9372298}
\begin{align}\label{Bellman}
Q^\mu(\mathbf{s}_t,\mathbf{a}_t)= \mathbb{E} \big[ r(\mathbf{s}_t,\mathbf{a}_t) + \gamma Q^\mu(\mathbf{s}_{t+1},\mathbf{a}_{t+1}) \big],
\end{align}
where $\mu$ represents the action policy function $\mathbf{S} \leftarrow \mathbf{A}$. The training of DRL agents aims for ascertaining the optimal action yielding the maximum Q value, as given by
\begin{align}
Q^*(\mathbf{s}_t,\mathbf{a}_t) = \mathbb{E} \big[ r(\mathbf{s}_t,\mathbf{a}_t) + \mathop{\max}_{\mathbf{a} \in \mathbf{A}}\gamma Q^*(\mathbf{s}_{t+1},\mathbf{a}_{t+1}) \big].
\end{align}

We consider a parameterized actor function $\mu(\mathbf{s}|\bm{\omega}^\mu)$ and a function approximator associated with a batch of parameters denoted by $\bm{\omega}^Q$. By sampling the aforementioned transition experiences in the memory of the replay buffer, the DRL agent can be trained by minimizing the loss function
\begin{align}\label{Loss}
L(\bm{\omega}^Q) = \frac{1}{e}\sum_{e}[y_t - Q(\mathbf{s}_t,\mathbf{a}_t|\bm{\omega}^Q_t)]^2,
\end{align}
where $e$ is the size of the sampled transitions. However, to avoid oscillations or divergence during the training process, $y_t$ has to be provided by the target network, which has the same structure as the training network, but associated with a deferred parameter update. Upon denoting the parameters of the target networks by $\bm{\omega}^{Q'}$, $y_t$ can be expressed as
\begin{align}\label{Y}
y_t = r_t(\mathbf{s}_t,\mathbf{a}_t)  + \gamma  Q'[\mathbf{s}_t,\mu'(\mathbf{s}_t|\bm{\omega}^{\mu'}_t)|\bm{\omega}^{Q'}_t].
\end{align}

According to the principle of the original DDPG algorithm \cite{lillicrap2015continuous}, the actor network is trained by the policy gradient calculated by the critic network as
\begin{align}\label{Actor}
&\nabla_{\bm{\omega}^\mu}J = \nonumber \\ &\frac{1}{e}\sum_{e}\nabla_\mathbf{a} Q(\mathbf{s}_e,\mathbf{a}_e|\bm{\omega}^Q)|_{\mathbf{s}_e=\mathbf{s}_t,\mathbf{a}_e=\mu(\mathbf{s}_t)}\nabla_{\bm{\omega}^\mu} \mu(\mathbf{s}_e|\bm{\omega}^\mu|_{\mathbf{s}_e=\mathbf{s}_t}).
\end{align}

\begin{figure*}[t!] \centering
\subfigure[Flow diagram of the conventional DDPG algorithm with output discretization.] {
 \label{Fig.DDPG}
\includegraphics[width=1.1\columnwidth]{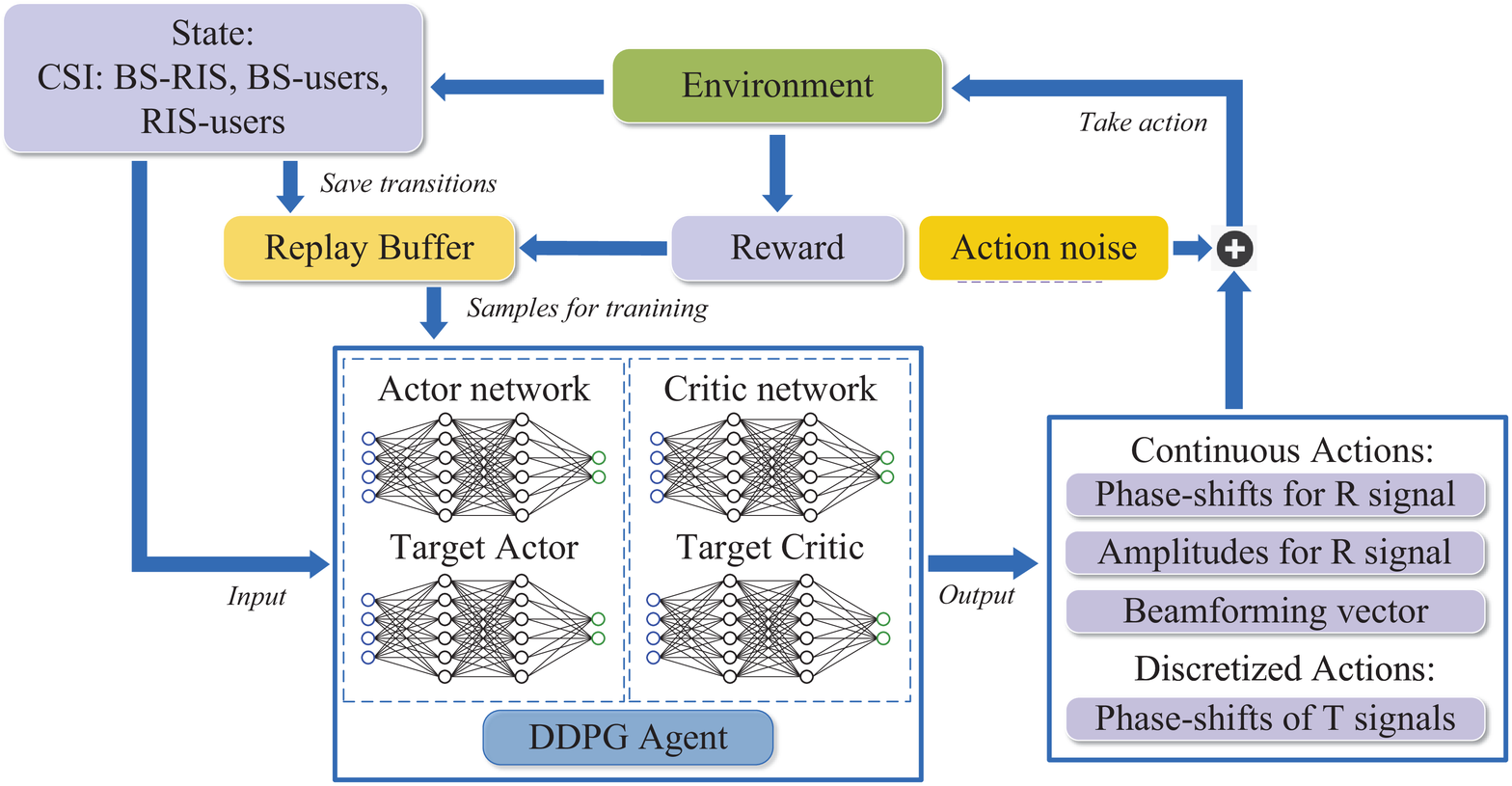}
}

\subfigure[Flow diagram of the proposed hybrid DDPG algorithm.] {
\label{Fig.HDDPG}
\includegraphics[width=1.1\columnwidth]{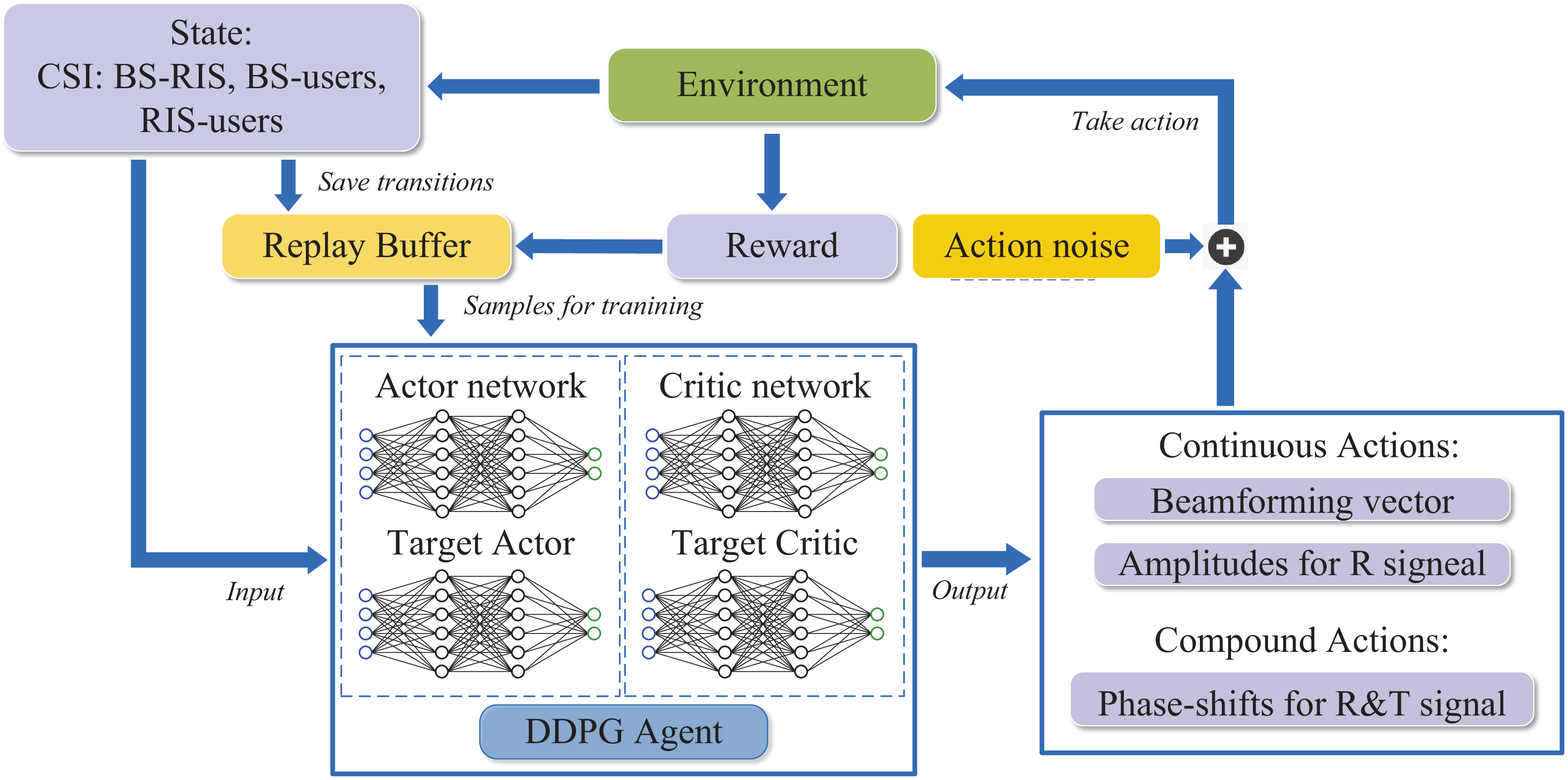}
}
\caption{Flow diagrams of the DDPG/hybrid DDPG algorithms.}
\label{Fig.DDPGs}
\end{figure*}

\begin{algorithm}
\caption{DDPG/Hybrid DDPG algorithm}
\label{DDPG}
\begin{algorithmic}[1]

        \STATE Initialize the environment and the agent with the actor network $\bm{\omega}^\mu$, critic network $\bm{\omega}^{Q}$, target actor network $\bm{\omega}^{\mu'}$, target critic network $\bm{\omega}^{Q'}$
        \FOR{each episode}
        \STATE Reinitialize the environment to $\mathbf{s}_{t=0}$
        \FOR{each step in $ t_0 \leq t \leq T $}
        \STATE Observe $\mathbf{s}_t$
        \STATE Choose $\mathbf{a}_t$ according to \eqref{A}
        \IF{DDPG algorithm}
            \STATE Discretize a part of $\mathbf{a}_t$ into $\mathbf{a}_{t,d}$
        \ENDIF
        \IF{Hybrid DDPG algorithm}
            \STATE Map $\mathbf{a}_t$ with $\bm{\Theta}_{\mathcal{R},t}$ and $\bm{\Theta}_{\mathcal{T},t}$
        \ENDIF
        \STATE Execute $\mathbf{a}_t$ in the environment
        \STATE Calculated the reward $r_t$ and observe the next state $\mathbf{s}_{t+1}$
        \STATE Record $e \{\mathbf{s}_t,\mathbf{a}_t,r_t,\mathbf{s}_{t+1}\}$ in memory buffer
        \STATE Random sample a batch of transection $e$ from memory buffer
        \STATE Calculate target according to \eqref{Y}
        \STATE Train critic network $Q(\mathbf{s}_t,\bm{\omega}^Q_t)$ with a gradient descent step \eqref{Loss}
        \STATE Train actor network ${\mu}(s_t,\bm{\omega}^{\mu}_t)$ with \eqref{Actor}
        \STATE Update the target networks $\bm{\omega}^\mathcal{\mu'}_t\leftarrow (1-\tau)\bm{\omega}^\mathcal{\mu'}_t + \tau\bm{\omega}^\mathcal{\mu}_t$, $\bm{\omega}^{Q'}_t\leftarrow (1-\tau)\bm{\omega}^{Q'}_t + \tau \bm{\omega}^Q_t$
        \STATE $\mathbf{s}_t \leftarrow \mathbf{s}_{t+1}$
        \ENDFOR
        \ENDFOR

\end{algorithmic}
\end{algorithm}

\subsection{Continuous-discrete actions and hybrid DDPG}\label{section:3B}

The actions designed for the DRL agent have to contain all optimized variables, resulting in the action space of $\mathbf{a}_t = \{ \mathbf{w},\bm{\Theta}_\mathcal{T},\bm{\Theta}_\mathcal{R},\bm{\beta}\}$. Among these optimization variables, $\mathbf{w}$ and $\bm{\beta}$ can be handled by a continuous control scheme to achieve precise control. Therefore, in this subsection we focus our attention on the discussion of $\bm{\Theta}_\mathcal{T}$ and $\bm{\Theta}_\mathcal{R}$.
As described above, due to the existence of the constraint \eqref{OPPF}, the  transmission and reflection phase-shifts cannot be adjusted independently by the STAR-RIS. For element $n$, assume $\beta_{n}\neq 0$ and $\theta_{\mathcal{R},n}$ is determined, in order to satisfy the constraint \eqref{OPPF}. In this context, it is no hard to discover that $\theta_{\mathcal{T},n}= \theta_{\mathcal{R},n}\pm \frac{\pi}{2}$. Therefore, regardless of whether the phase-shift $\theta_{\mathcal{R},n}$ or $\theta_{\mathcal{T},n}$ is selected to be continuously controlled, the phase control of the other one is no longer a continuous control problem, but a binary selection problem. Therefore, \eqref{OPP} requires continuous control for $\mathbf{w},\bm{\Theta}_\mathcal{R},\bm{\beta}$ and discrete control for $\bm{\Theta}_\mathcal{T}$, which lead to a hybrid action space associated with continuous and discrete actions. The continuous sub-space and the discrete sub-space are denoted by $\mathbf{a}^c = (a^c_1,a^c_2...a^c_n...)$ and $\mathbf{a}^d = (a^d_1,a^d_2...a^d_n...)$\footnote{For the convenience of presentation, we assume that $\bm{\Theta}_\mathcal{R} \subset a^c$  and $\bm{\Theta}_\mathcal{T} \subset a^d$.}, respectively.

\begin{figure}[t!]
    \begin{center}
        \includegraphics[width=0.4\textwidth]{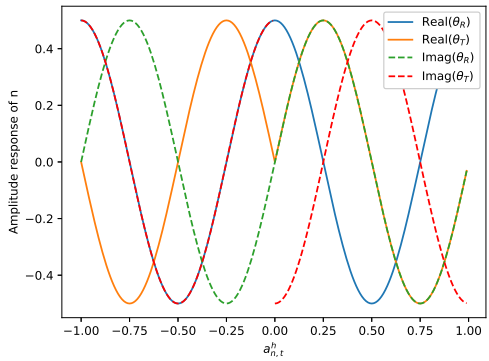}
        \caption{Amplitude response over normalized action output of hybrid DDPG algorithm for STAR-RIS ($\beta=0.5$).}
        \label{Fig.actionresponse}
    \end{center}
\end{figure}

\subsubsection{Action Space for the DDPG Algorithm}

If the DDPG algorithm is applied for solving the optimization problem associated with the hybrid action space, a direct and intuitive approach is to discretize a part of the continuous output of the DDPG algorithm as presented in Fig.~\ref{Fig.DDPG} and \textbf{Algorithm \ref{DDPG}}. Including the action noise, a classic action policy is given for the DDPG algorithm by
\begin{align}\label{DDPGAction}
\mathbf{a}_t = \mu(\mathbf{s}_t|\bm{\omega}^\mu_t)+\mathcal{N}_\text{OU}(0,\xi),
\end{align}
where $\mathcal{N}_\text{OU}(0,\xi)$ represents the zero mean Ornstein Uhlenbeck (OU) noise \cite{9312988} that follows $\mathcal{N}_\text{OU} \sim OU(0,\xi)$, and $ \xi$ is the volatility of the OU noise.

Thus, the continuous action space can be formulated as
\begin{align}\label{A}
\mathbf{a}_t^c = \{ \mathbf{a}_t^\mathbf{w}, \mathbf{a}_t^{\bm{\Theta}_\mathcal{R}} , \mathbf{a}_t^{\bm{\beta}} \}.
\end{align}

Then, the normalized output of the actor network may be decoded into executable actions for our communication environment by following the actions of
\begin{align}\label{docoding1}
&\mathbf{w} \leftarrow \mathbf{a}_t^\mathbf{w},\\ \label{docoding2}
&\bm{\Theta}_\mathcal{R} \leftarrow \mathbf{a}_t^{\bm{\Theta}_\mathcal{R}}, \\ \label{docoding3}
&\bm{\beta} \leftarrow \mathbf{a}_t^{\bm{\beta}}.
\end{align}

Since the mapping of $\mathbf{a}_t^c$ and the continuous actions are trivial, it is not necessary to discuss them in detail. On the other hand, for a specific STAR element $n$, $\theta_{\mathcal{T},n}$ can be obtained by the binary discretized $a_{n,t}^d$
\begin{align}\label{docoding4}
\theta_{\mathcal{T},n,t} =
\begin{cases}
\theta_{\mathcal{R},n,t}+\frac{\pi}{2}, & a_{n,t}^d >0,\\
\theta_{\mathcal{R},n,t}-\frac{\pi}{2}, & a_{n,t}^d \le 0.
\end{cases}
\end{align}

\subsubsection{Action Space for our Hybrid DDPG Algorithm}

In order to deal with the high-dimensional continuous and discrete action components, specifically for the STAR-RIS scenario, we propose a hybrid DDPG algorithm. By exploiting $\bm{\Theta}_\mathcal{R}$ and $\bm{\Theta}_\mathcal{T}$ have the same dimension for the STAR-RIS, we design the codebook between the actor outputs and the phase-shifts to achieve hybrid control. The hybrid action policy for the STAR-RIS at time $t$ can be expressed by
\begin{align}\label{HDDPGAction}
\mu(\mathbf{s}_t|\bm{\omega}^\mu_t) &= \mu^c(\mathbf{s}_t|\bm{\omega}^\mu_t)\mu^d(\mathbf{s}_t|\bm{\omega}^\mu_t) \nonumber \\ &=\prod_{n=1}^{N}\mu^c(s_{n,t}|\bm{\omega}^\mu_t)  \prod_{n=1}^{N}\mu^d(s_{n,t}|\bm{\omega}^\mu_t).
\end{align}

Specifically, for the STAR element $n$, the values of $\theta_{\mathcal{R},n,t}$ and $\theta_{\mathcal{T},n,t}$ are compounded and given in a normalized output as
\begin{align}\label{HDDPGAction2}
\mu(s_{n,t}|\bm{\omega}^\mu_t) =\mu^c(s_{n,t}|\bm{\omega}^\mu_t) \mu^d(s_{n,t}|\bm{\omega}^\mu_t).
\end{align}

In the face of the action noise similar to \eqref{DDPGAction}, the hybrid action $a^h_{n,t}$ can be obtained from \eqref{HDDPGAction2}. Then, the normalized action $a^h_{n,t}$ has to be mapped to $\theta_{\mathcal{R},n,t}$ and $\theta_{\mathcal{R},n,t}$ as
\begin{align}
\theta_{n,\mathcal{R},t} = 2\pi a^h_{n,t},
\end{align}
\begin{align}
\theta_{n,\mathcal{T},t} =
\begin{cases}
\theta_{n,\mathcal{R},t}+\frac{\pi}{2}, & a^h_{n,t} >0,\\
\theta_{n,\mathcal{R},t}-\frac{\pi}{2}, & a^h_{n,t} \le 0.
\end{cases}
\end{align}

Since the normalized action $a^h_{n,t}$ is in the interval $[-1,1]$, $\theta_{n,\mathcal{R},t}$ and $\theta_\mathcal{T}$ have to be mapped with $a^h_{n,t}$. Furthermore, $\theta_{n,\mathcal{R},t}$ vs $a^h_{n,t}$ is modelled by a periodic linear function and $\theta_\mathcal{T}$ vs $a^h_{n,t}$ by a piecewise linear function. The amplitude response $a^h_{n,t}$ of a s single STAR element is plotted in Fig. \ref{Fig.actionresponse} for $\beta=0.5$.  Since STAR-RIS is generally equipped with a large number of STAR elements, in this case, the hybrid DDPG algorithm has a significantly smaller action dimension than the conventional DDPG algorithm, since we have $|\mathbf{a}^h_{t}| = \frac{|\mathbf{a}^c_{t}| + |\mathbf{a}^d_{t}|}{2}$.

\begin{remark}
In some existing hybrid DRL schemes, the agent can only output a single action for the discrete action space, which is a disadvantage for problems associated with high-dimensional discrete action spaces of multi-antenna or multi-element-RIS scenarios. By contrast, the proposed hybrid scheme is eminently suitable for the high-dimensional hybrid action spaces, where the discrete action dimension is no larger than the continuous action dimension.
\end{remark}

\subsection{Reward Function}

The plain DDPG algorithm and the hybrid DDPG algorithm have identical reward functions. In order to ensure that the agent meets the users' QoS constraint, each satisfied $\mathcal{T}$ and $\mathcal{R}$ user contributes a distinguishable positive reward. Meanwhile, to minimize the power consumption, the sum power cost associated with $\mathbf{w}$ results in a negative reward. Thus, the reward function at TS $t$ can be formulated as
\begin{equation}\label{R}
r_t= \sum_{k=1,R_{k,t}>R_c}^{K} \dot{r} - \sum_{t=1}^{T} \sum_{k=1}^{K} \hat{r} \parallel \mathbf{w}^2_{k,t}\parallel,
\end{equation}
where the constant coefficient $\dot{r}$ represents the reward gleaned for satisfying the data rate requirement per user, and the coefficient $\hat{r}$ cost (negative reward) for power consumption. It is worth noting that, to guarantee the primary goal of the agent is satisfying the QoS requirement of each and every user rather than saving energy, the coefficients have to satisfy $ \dot{r} > \sum_{k=1}^{K} \hat{r} P_\text{max}$.

\subsection{Neural Network Structure}\label{section:3C}

In order to ensure accurate fitting, the structure and scale of the (target) actor network and the (target) critic network have to be selected appropriately. The actor networks include the input layer, batch normalization (BN) layer, and activation layer(s) with a 'relu' function in turn. Moreover, 'tanh' is assigned as the activation function of the output layer. To ensure a valid range for the input values of the output layer, another BN layer has to be employed above the output layer. The critic networks consist of an input layer, BN layer, concatenate layer, and activation layers.
The size of the hidden layers has to be determined according to the state and action dimension, which is dependent both on the number of antennas $M$ and the number of STAR elements $N$. It is also worth noting that due to the difference in output dimensions between the DDPG scheme and the hybrid DDPG scheme, the appropriate size of the hidden layer may vary, especially when a large number of STAR elements is considered.

\section{Solution II: Joint DDPG-DQN Algorithm}\label{section:4}

The hybrid DDPG algorithm aims for covering the action space $\mathbf{a}^c$ and $\mathbf{a}^d$  with the aid of a compound action mapping. By contrast, in this section, we explore another promising option, namely that of employing two agents for covering the action spaces $\mathbf{a}^c$ and $\mathbf{a}^d$, respectively. Thus, a DDPG algorithm is harnessed for optimizing $\mathbf{a}^c = \{ \mathbf{a}_t^\mathbf{w}, \mathbf{a}_t^{\bm{\Theta}_\mathcal{R}} , \mathbf{a}_t^{\bm{\beta}}\}$, and a DQN algorithm is responsible for giving $\mathbf{a}^d = \mathbf{a}_t^{\bm{\Theta}_\mathcal{T}}$. DDPG and DQN algorithms can be regarded either as a joint agent, or as a pair of cooperative agents. For the convenience of the presentation, we regard the DDPG agent and the DQN agent of the joint DDPG-DQN algorithm as a pair of components, but they have to be installed on the same device in practice.

\subsection{MDP for Joint DDPG-DQN}\label{section:4A}

The DDPG and DQN algorithms are capable of optimizing independent continuous and discrete problems \cite{8103164}. Nonetheless, how to harness them in the interest of joint optimization results is an open conundrum. If two agents adopt a parallel relationship at $\mathbf{s}_t$ to output $\mathbf{a}^c_t$ and $\mathbf{a}^d_t$, then according to the classic MDP model, the execution of $\mathbf{a}_t$ will result in reward $r_t$. However, both agents need corresponding rewards for estimating the action value of $\mathbf{a}^c_t$ and $\mathbf{a}^d_t$ independently. The dilemma is that in the scenario considered, $r_t$ is the result of the combined action of $\mathbf{a}^c_t, \mathbf{a}^d_t$ and cannot be split into $r^c_t, r^d_t$. In other words, from the perspective of the wireless network, the user-side SINR of the STAR-RIS network is jointly determined by active beamforming, reflection beamforming, and transmission beamforming. As a consequence, we cannot solely and unambiguously attribute any gain or loss. If $r_t$ is regarded as a common rewards for both two agents, the agents will authenticate the MDP transitions as $\{\mathbf{s}_{t},\mathbf{a}^c_{t},r_{t},\mathbf{s}_{t+1}\}$ and $\{\mathbf{s}_{t},\mathbf{a}^d_{t},r_{t},\mathbf{s}_{t+1}\}$. Unfortunately, these transitions are not correct, as in fact the correct transitions have to be formulated as $\{\mathbf{s}_{t},\mathbf{a}_{t},r_{t},\mathbf{s}_{t+1}\}$ or $\{\mathbf{s}_{t},\mathbf{a}^c_{t},r^c_{t},\mathbf{s}^c_{t+1}\}, \{\mathbf{s}_{t},\mathbf{a}^d_{t},r^d_{t},\mathbf{s}^d_{t+1}\}$. In order to resolve this dilemma, inspired by \cite{li2021joint}, we artificially add an inner environment to formulate the MDPs for a pair of agents.

\begin{figure*}[t!]
    \begin{center}
        \includegraphics[width=0.6\textwidth]{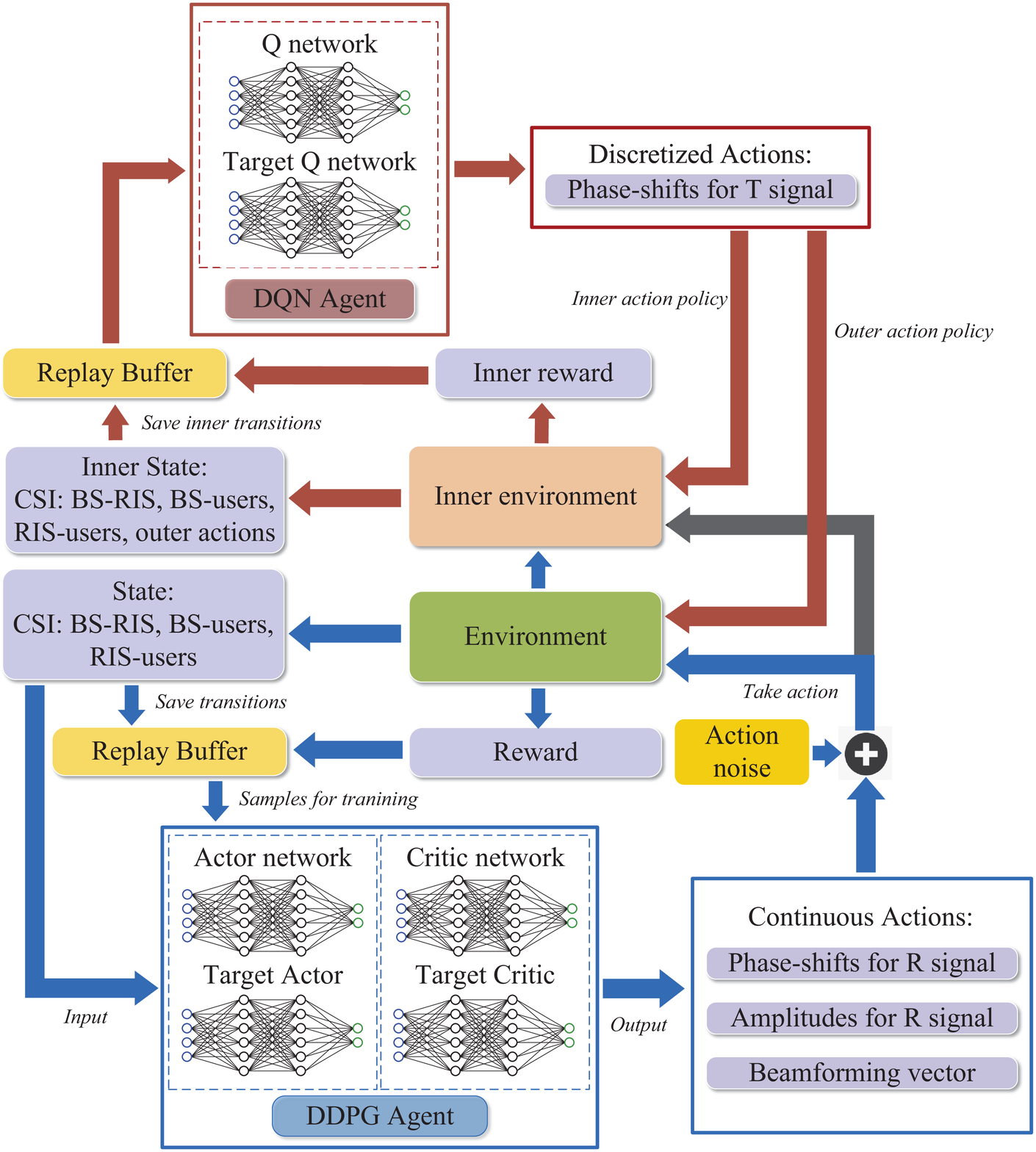}
        \caption{Flow diagram of the proposed DDPG-DQN algorithm.}
        \label{Fig.DDPGDQN}
    \end{center}
\end{figure*}

As shown in Fig. \ref{Fig.DDPGDQN}, an extra environment is derived from the original environment, and the environments refer to the 'inner environment' and 'outer environment' to distinguish them. Thus, the DDPG agent interacts with the outer environment and the DQN agent interacts with the inner environment. The outer environment represents the STAR-RIS assisted wireless network in the reality, the inner environment is only a fictitious environment that contains the knowledge of the outer environment for the agent's training. From the perspective of the DQN agent, the inner environment can be regarded as a collection of the outer environment and the DDPG agent. Based on this framework, for the DQN agent and the inner environment, the MDP can be formulated as $\{\mathbf{s}_{i,t},\mathbf{a}^d_{i,t},r_{i,t},\mathbf{s}_{i,t+1}\}$ and for the outer environment as $\{\mathbf{s}_{o,t},\mathbf{a}^c_{o,t},r_{o,t},\mathbf{s}_{o,t+1}\}$. In the remainder of this section, we discuss the details of the DQN and DDPG agents, respectively.

\begin{algorithm}
\caption{Joint DDPG-DQN algorithm}
\label{DDPGDQN}
\begin{algorithmic}[1]

        \STATE Initialize the outer environment and the DDPG agent with the actor network $\bm{\omega}^\mu$, critic network $\bm{\omega}^{Q}$, target actor network $\bm{\omega}^{\mu'}$, target critic network $\bm{\omega}^{Q'}$
        \STATE Initialize the DQN agent with the deep Q network $\bm{\omega}^\mathcal{Q}$ and target Q network $\bm{\omega}^\mathcal{Q'}$
        \FOR{each episode}
        \STATE Reinitialize the outer environment to $\mathbf{s}_{t=0}$
        \FOR{each step in $ t_0 \leq t \leq T $}
        \STATE Observing $\mathbf{s}_{o,t}$ and DDPG agent choose $\mathbf{a}^c_{o,t}$ according to \eqref{A}
        \STATE DQN agent choose inner action $\mathbf{a}^d_{i,t}$ with \eqref{actionpolicyinner}
        \STATE Execute action $\mathbf{a}^d_{i,t}$ in the inner environment
        \STATE Calculated the reward $r_{i,t}$ and observe the next state $\mathbf{s}_{i,t+1}$
        \STATE Record $e \{\mathbf{s}_{i,t},\mathbf{a}^d_{i,t},r_{i,t},\mathbf{s}_{i,t+1}\}$
        \STATE Sample random transitions of $e$ from DQN memory
        \STATE Train $\omega^\mathcal{Q}$ with \eqref{LossDQN}
        \STATE DQN agent choose outer action $\mathbf{a}^d_{o,t}$ with \eqref{actionpolicyouter}
        \STATE Execute $\mathbf{a}^c_{o,t}$ and $\mathbf{a}^d_{o,t}$ in the outer environment
        \STATE Calculated the reward $r_{o,t}$ and observe the next state $\mathbf{s}_{r,t+1}$
        \STATE Record $e \{\mathbf{s}_{o,t},\mathbf{a}_{o,t},r_{o,t},\mathbf{s}_{o,t+1}\}$ in DDPG memory buffer
        \STATE Random sample a batch of transection $e$ from memory buffer
        \STATE Calculate target according to \eqref{Y}
        \STATE Train critic network $Q(\mathbf{s}_t,\bm{\omega}^Q_t)$ with a gradient descent step \eqref{Loss}
        \STATE Train actor network ${\mu}(\mathbf{s}_t,\bm{\omega}^{\mu}_t)$ with \eqref{Actor}
        \STATE Update the target networks: $\bm{\omega}^\mathcal{\mu'}_t\leftarrow (1-\tau)\bm{\omega}^\mathcal{\mu'}_t + \tau\bm{\omega}^\mathcal{\mu}_t$, \\ $\bm{\omega}^{Q'}_t\leftarrow (1-\tau)\bm{\omega}^{Q'}_t + \tau \bm{\omega}^Q_t$, $\bm{\omega}^\mathcal{Q'}_t\leftarrow (1-\tau)\bm{\omega}^\mathcal{Q'}_t + \tau \bm{\omega}^\mathcal{Q}_t$
        \STATE $\mathbf{s}_{o,t} \leftarrow \mathbf{s}_{o,t+1}$
        \ENDFOR
        \ENDFOR

\end{algorithmic}
\end{algorithm}

\subsection{Inner Environment and the DQN agent}\label{section:4B}
\subsubsection{DQN Training}

As a value-based RL algorithm, the DQN algorithm identifies the action values as \eqref{Bellman}, since the DQN agent also aims for the maximum long-term reward. In the inner environment, the Q function is given by
\begin{align}\label{BellmanDQN}
Q(\mathbf{s}_{i,t},\mathbf{a}^d_{i,t})= \mathbb{E} \big[r(\mathbf{s}_{i,t},\mathbf{a}^d_{i,t}) + \gamma Q(\mathbf{s}_{i,t+1},\mathbf{a}_{i,t+1}) \big].
\end{align}

In the training process, the Q value for the actions has to be updated in each step by following
\begin{align}\label{Qupdate}
Q_{t+1}(\mathbf{s}_{i,t},\mathbf{a}^d_{i,t}) &\leftarrow (1-\alpha)Q_t(\mathbf{s}_{i,t},\mathbf{a}^d_{i,t})\nonumber\\ &+ \alpha[r_{i,t} + \gamma \max Q_{t}(\mathbf{s}_{i,t+1},\mathbf{a}^d_{i,t+1})],
\end{align}
where $\alpha$ represents the learning rate ( $0<\alpha\leq1$). In the DQN algorithm, the Q function is approximated by a DNN having the parameter vector $\bm{\omega}^\mathcal{Q}$, and we have $Q(\mathbf{s}_{i,t},\mathbf{a}^d_{i,t}) \approx Q(\mathbf{s}_{i,t},\mathbf{a}^d_{i,t}|\omega^\mathcal{Q})$. In order to accurately fit the Q function, the DNN has to be appreciatively trained. Similar to the DDPG algorithm, the transitions of the DQN agent are stored in the DQN memory buffer and we can adopt the memory replay technique for training the DNN by the following loss function
\begin{align}\label{LossDQN}
&L(\omega^\mathcal{Q}) = \frac{1}{e}\sum_{e}\big[r_{i,t}(\mathbf{s}_{i,t},\mathbf{a}^d_{i,t})+  \nonumber\\ &\gamma\!\!  \mathop{\max}_{\mathbf{a}_{i,t+1} \in \mathbf{A}^d} Q'(\mathbf{s}_{i,t+1},\mathbf{a}_{i,t+1}|\omega^{\mathcal{Q}'}_t) - Q(\mathbf{s}_{i,t},\mathbf{a}_{i,t}|\omega^\mathcal{Q}_t)\big]^2.
\end{align}

\subsubsection{Inner State and Action Space}

As designed above, the DQN agent regards the DDPG agent as a part of the inner environment. Therefore, the state of the inner environment consists of two parts, including the state inherited from the outer environment and the actions $\mathbf{a}^c_t$ of the DDPG agent. Thus the inner state can be formulated as
\begin{align}
\mathbf{s}_{i,t} =\{\bm{H}_{b,r,t},\bm{H}_{b,\mathcal{R},t},\bm{H}_{b,\mathcal{T},t},\bm{H}_{r,\mathcal{R},t},\bm{H}_{r,\mathcal{T},t}, \mathbf{a}^c_{o,t} \}.
\end{align}

The action space of the DQN agent is $\mathbf{a}^d_{i,t}=\bm{\Theta}_\mathcal{T}$, since the DQN agent is only responsible for the discrete actions. The state space and action indicate a pair of main facts. Primarily, the decision of the DDPG agent has to be known to the DQN agent, when it is interacting with the inner environment.  Additionally, it also suggests that the DQN agent has to 'interrupt' the interactions of the DDPG agent. Specifically, as shown in \textbf{Algorithm \ref{DDPGDQN}}, the timing of activating the inner environment and the DQN agent is after the action selection of the DDPG agent, but before the outer environment's execution of the action, since the transition of the outer environment relies on $\mathbf{a}_{o,t}=\{\mathbf{a}^c_{o,t}, \mathbf{a}^d_{o,t}\}$.

\subsubsection{Action Policy}

For conventional DQN agents, the $\epsilon-\text{greedy}$ action policy constitutes an efficient technique of carefully balancing exploration and exploitation. To elaborate, the $\epsilon-\text{greedy}$ policy authorizes the DQN agent to choose a random action with the probability of $\epsilon$, and the optimal action with a probability of $1-\epsilon$. However, it is worth noting that in the joint DDPG-DQN algorithm, the DQN agent has to output a pair of actions one for the inner and one for the outer environment. Action $\mathbf{a}^d_{i,t}$ for the inner environment is adopted to form the inner transitions for the training process of the DQN agent. On the other hand, the action $\mathbf{a}^d_{o,t}$ produced by the DQN for the outer environment has to assist the DDPG agent. Therefore, once the $\epsilon-\text{greedy}$ action policy is adopted in the outer environment, it may impose interference on $r_{o,t}$ and lead to an estimation error on the Q value of $\mathbf{a}^c_{o,t}$. Thus, the action policy of the DQN agent can be formulated as

\begin{equation}\label{actionpolicyinner}
\mathbf{a}^d_{i,t}=
\begin{cases}
\text{random action}, &  \epsilon, \\
\mathop{\arg\max}\limits_{\mathbf{a}_{i,t} \in \mathbf{A}^d} Q(\mathbf{s}_{i,t+1},\mathbf{a}_{i,t}|\bm{\omega}^{Q}_t) ,&  1-\epsilon,
\end{cases}
\end{equation}

\begin{equation}\label{actionpolicyouter}
\mathbf{a}^d_{o,t}= \mathop{\arg\max}\limits_{\mathbf{a}_{i,t} \in \mathbf{A}^d} Q(\mathbf{s}_{o,t},\mathbf{a}_{o,t}|\bm{\omega}^{Q}_t).
\end{equation}

\begin{remark}\label{Remark2}
In the joint DDPG-DQN algorithm, the DQN is trained in the inner environment. The DQN agent produces actions for the outer environment, which may be viewed as a service instead of training for the DQN. Therefore, we choose the $\epsilon-\text{greedy}$ action policy in the inner environment to train the DQN agent and the optimal action policy in the outer environment to obtain optimal actions.
\end{remark}

\subsubsection{Reward Function}

Since the reward function depends on the optimization goal and on the constraints, similarly to \eqref{R}, the reward function of the DQN agent is also given by

\begin{equation}\label{RDQN}
r_{i,t}= \sum_{k=1,R_{k,t}>R_c}^{K} \dot{r} - \sum_{t=1}^{T} \sum_{k=1}^{K} \hat{r} \parallel \mathbf{w}^2_{k,t}\parallel,
\end{equation}
where the constant settings have to be consistent with the corresponding discussion of Sub-section \ref{section:3C}.

\subsection{Outer Environment and the DDPG agent}\label{section:4C}

The principle and algorithm's flow of the outer environment and the DDPG agent are the same as in Section \ref{section:3}, thus we do not repeat the principle and training process here but highlight the difference. For the DDPG agent, state $\mathbf{s}_{o,t}$ is the same as \eqref{S}, and the reward is $r_{o,t}= \sum_{k=1,R_{k,t}>R_c}^{K} \dot{r} - \sum_{t=1}^{T} \sum_{k=1}^{K} \hat{r} \parallel \mathbf{w}_k^2\parallel$, but the difference is that the action space is reduced to $\mathbf{a}^c_{o,t}=\{ \mathbf{a}_t^\mathbf{w}, \mathbf{a}_t^{\bm{\Theta}_\mathcal{R}} , \mathbf{a}_t^{\bm{\beta}} \}$. Hence, for any TS $t$, the optimized value $\{ \mathbf{w}_t,\bm{\Theta}_{\mathcal{R},t},\bm{\beta}_t\}$ can be completely covered by $\mathbf{a}^c_{o,t}$ using the mapping approach described in \eqref{docoding1} \eqref{docoding2} and \eqref{docoding3}.

In the joint DDPG-DQN algorithm, since both agents are employed and each has its own DNNs, their optimal DNN scales are likely to be smaller than these of the hybrid DDPG algorithm. For the DDPG agent in the joint algorithm, we adopt the same structure as for the DNNs of the hybrid DDPG algorithm, not only as its optimal experimental structure, but also to ensure the fairness of the performance comparison.

\subsection{Discussions}\label{section:4D}

Although facing an identical formulated problem, Solution I and Solution II revealed two disparate measures. The hybrid DDPG approach aimed for achieving hybrid control via carefully tailored mapping function without any additional DNN or other processes. The advantage of this scheme is that its complexity is not increased. By contrast, since it has a lower output dimension than the DDPG agent, i.e. $ |\mathbf{a}^h_{n}| < |\mathbf{a}^c_{n}|+|\mathbf{a}^d_{n}|$, it can employ a DNN with fewer trainable parameters, thereby reducing the complexity of the training process. The joint DDPG-DQN algorithm employs a pair of agents having different capabilities for jointly solving the problem having a hybrid action space. Since an extra DQN agent has to be harnessed, given the complexity of the DDPG and the DQN algorithm \cite{qiu2019deep}, the total complexity becomes significantly higher than that of the hybrid DDPG scheme.

The complexity of the DDPG algorithm depends on the specification of the employed DNN. Assuming that an actor network having $I$ layers is employed, while each layer contains $\bm{\omega}^\mu_i$ nodes, the complexity of propagation is given by $\sum \limits_{i=0}^{I}\bm{\omega}^\mu_i \bm{\omega}^\mu_{i+1}$. Given the number of nodes in the actor network as $\bm{\omega}^\mu_\text{b}$ for BN layers, $\bm{\omega}^\mu_\text{r}$ for 'relu' layers and $\bm{\omega}^\mu_\text{t}$ for 'tanh' layers, according to \cite{qiu2019deep}, the required number of floating-point operations is given by $ 5 \bm{\omega}^\mu_\text{b} + \bm{\omega}^\mu_\text{r} + 6 \bm{\omega}^\mu_\text{t}$. Applying the same theory for the critic networks, the complexity of a single prediction and training step can be formulated by $\mathcal{O}(\sum \limits_{i=0}^{I}\bm{\omega}^\mu_i \bm{\omega}^\mu_{i+1} + \sum \limits_{i=0}^{I}\bm{\omega}^Q_i \bm{\omega}^Q_{i+1} + 5 \bm{\omega}^\mu_\text{b} + \bm{\omega}^\mu_\text{r} + 6 \bm{\omega}^\mu_\text{t} + 5 \bm{\omega}^Q_\text{b} + \bm{\omega}^Q_\text{r} + 6 \bm{\omega}^Q_\text{t})$. Since we usually have $\bm{\omega}_{i} \gg 5$, the computational complexity can be approximated by $\mathcal{O}(\sum \limits_{i=0}^{I}\bm{\omega}^\mu_i \bm{\omega}^\mu_{i+1} + \sum \limits_{i=0}^{I}\bm{\omega}^Q_i\bm{\omega}^Q_{i+1})$.  As mentioned, since the hybrid DDPG algorithm has smaller output and DNN scale, we have $\bm{\omega}^{\mu,h}_i < \bm{\omega}^\mu_i$. Thus, the hybrid DDPG algorithm has lower complexity than the conventional DDPG algorithm.

The complexity of the joint DDPG-DQN algorithm is partially due to the DDPG agent and partially from the DQN agent. The complexity of the DQN agent is given by $\mathcal{O}(3\sum \limits_{i=0}^{I}\bm{\omega}^\mathcal{Q}_i\bm{\omega}^\mathcal{Q}_{i+1})$ as suggested in \cite{9354068}. The overall complexity of the joint DDPG-DQN algorithm is $\mathcal{O}(\sum \limits_{i=0}^{I}\bm{\omega}^\mu_i \bm{\omega}^\mu_{i+1} + \sum \limits_{i=0}^{I}\bm{\omega}^Q_i\bm{\omega}^Q_{i+1} + 3\sum \limits_{i=0}^{I}\bm{\omega}^\mathcal{Q}_i\bm{\omega}^\mathcal{Q}_{i+1})$. Therefore, the joint DDPG-DQN algorithm only has an increased complexity compared to the hybrid DDPG algorithm if the DQN agent has a significantly smaller $\bm{\omega}^\mathcal{Q}_i$.

\section{Numerical Results and Analysis}\label{section:5}

Let us now consider the performance of the proposed approaches.  The performance of the STAR-RIS is compared to that of the reflecting-only RIS and double spliced RIS. The reflecting-only RIS can only serve $\mathcal{R}$ users and fails to provide signal enhancements for the $\mathcal{T}$ users. As for the other benchmark, the double spliced RIS is formed by splicing a pair of RISs facing in opposite directions, where the total number of elements is set to $N$. This also ensures the fairness of comparisons. The double spliced RIS can also be equivalently regarded as a STAR-RIS employing the 'Mode Switching' scheme \cite{mu2021simultaneously}, but the proportion of transmission elements and reflection elements is fixed to 1. In terms of optimization algorithms, the DDPG algorithm having a partial discrete action space is employed as the baseline. The performance of the hybrid DDPG algorithm and DDPG-DQN algorithm is compared.

As for the simulation parameters, the (STAR) RIS located 4km away from the BS, where the $\mathcal{T}$ users and $\mathcal{R}$ users are randomly distributed on both sides of the (STAR) RIS. The wireless channel model was introduced in Section \ref{section:2}. The simulated transmission process is 30 seconds and we assume that the block fading envelope of the channel varies once per second. For the intelligent agent, we employ the 'Adam' optimizer for both the hybrid DDPG and for the DDPG-DQN algorithms. The default learning rates for all agents, including the learning rate for the DQN, the actor/critic learning rates for the DDPG agent are set to be $3\times10^{-4}$. For the DDPG agent, the actor network has a single activation layer, and the critic network has two activation layers. As for the DQN agent, the deep Q network has 1-2 activation layer(s). The activation function of the hidden layers is 'relu' in both algorithms. As mentioned, the size of the activation layers has to be determined by the complexity of the communication system. Empirically, in the majority of realizations, we equip the actor and critic network with hidden layer(s) containing 256-512 neuron nodes, and the DQN agent employs hidden layer(s) with 200-300 neuron nodes. The default parameters used in the simulations are listed in Table \ref{SP}.

\begin{table*}[t!]
 \caption{Default Parameters}\label{SP}
 \centering
 \footnotesize
 \renewcommand\arraystretch{1.5}
 \begin{tabular}{|c|c|c|c|c|c|}
  \hline
  Parameter & Description & Value &   Parameter & Description & Value \\
  \hline
  $M$ & antenna number & 4 & $N$ & (STAR) element number  & 12  \\
  \hline
  $f_\text{c}$ & carrier frequency & 5GHz & $B$ & bandwidth  & 1 MHz  \\
  \hline
  $K$  & number of users & 4 & $P_\text{max}$ & maximum power per antenna & 29 dBm \\
  \hline
  ${K_{AR},K_{RU}}$& Rician factors & 3dB & $\sigma$& noise power density & -95.2 dBm/MHz \\
  \hline
  $r$ & replay buffer size & 10000 & $\gamma$ & discount factor & 1\\
  \hline
  $e_\text{DDPG}$ & batch size & 32 samples & $e_\text{DQN}$ & batch size & 32 samples\\
  \hline
  $\tau_\text{DDPG}$ & target update rate & 0.002  & $\tau_\text{DQN}$ & target update rate & 0.003\\
  \hline
 \end{tabular}
\end{table*}

\begin{figure}[t!]
    \begin{center}
    \setlength{\belowcaptionskip}{-0.3cm}
        \includegraphics[width=0.45\textwidth]{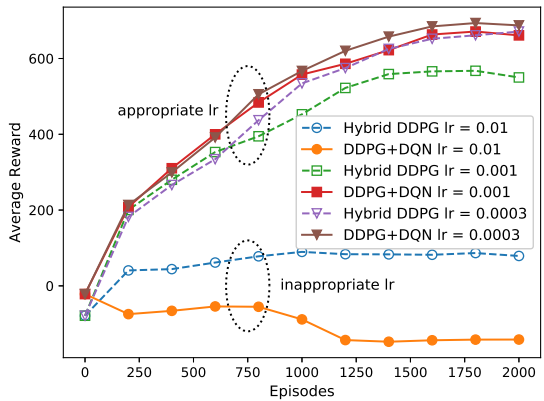}
        \caption{Obtained reward of hybrid DDPG algorithm and DDPG-DQN algorithm with different learning rate}
        \label{Fig.reward}
    \end{center}
\end{figure}

Fig. \ref{Fig.reward} presents the reward obtained by the proposed algorithms having different learning rates. Based on the rewards exhibited by the different learning rates, we can draw two main conclusions. On the one hand, both the hybrid DDPG and the DDPG-DQN schemes converge if the appropriate learning rate is selected. It has to be clarified that, in general, a higher learning rate leads to faster convergence. However, the curves $\text{lr}=0.001$ and $\text{lr}=0.0003$ of Fig. \ref{Fig.reward} exhibit similar convergence rates, since we invoked decaying action noise and exploration rate in the (hybrid) DDPG and DQN agents, which affects the reward obtained. On the flip side, the joint DDPG-DQN approach has an excellent capability of obtaining rewards. According to \eqref{R}, the rewards obtained by the agents represent the overall performance of the wireless links, which translate into superior user satisfaction or reduced energy consumption.

\begin{figure}[t!]
    \begin{center}
    \setlength{\belowcaptionskip}{-0.3cm}
        \includegraphics[width=0.45\textwidth]{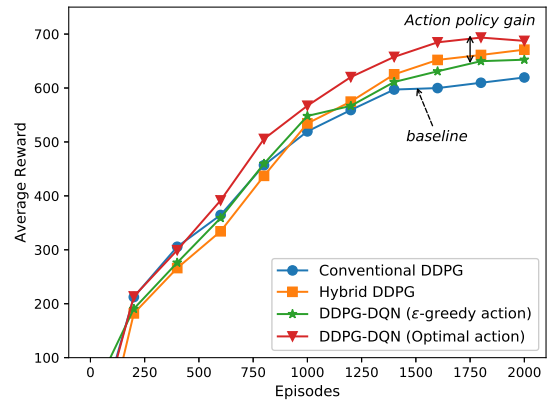}
        \caption{Performances of different algorithms for STAR-RIS }
        \label{Fig.algorithmreward}
    \end{center}
\end{figure}

The rewards obtained by the different algorithms are plotted in Fig. \ref{Fig.algorithmreward}. After a training process, the conventional DDPG algorithm used as the baseline has achieved an average reward of about 590, which is inferior to the proposed algorithms. The DDPG-DQN algorithm having optimal action output achieved a slight advantage of about $3-6\%$ compared to that of the hybrid DDPG algorithm, but the price of this performance gain is that the DDQP-DQN algorithm has higher complexity, since two agents are employed. Although the hybrid DDPG scheme is slightly inferior in terms of optimality, it provides a low-complexity rapidly converging solution. Observing the two branches of the DDPG-DQN scheme, it can be observed that compared to the scheme in \cite{li2021joint}, the separated action policy of inner and outer environments has achieved superior performance, which supports the arguments in \textbf{Remark \ref{Remark2}}.


\begin{figure}[t!]
    \begin{center}
    \setlength{\belowcaptionskip}{-0.3cm}
        \includegraphics[width=0.45\textwidth]{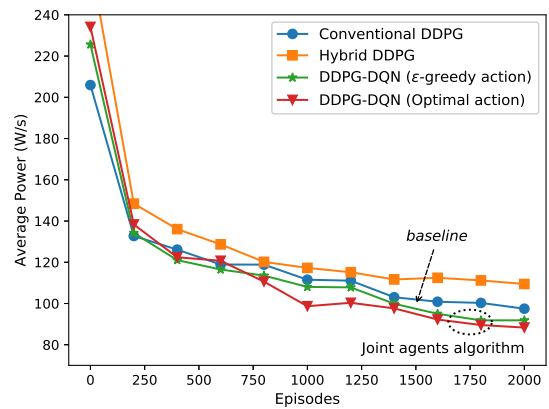}
        \caption{Power consumption of different algorithms for STAR-RIS}
        \label{Fig.algorithmpower}
    \end{center}
\end{figure}

Fig. \ref{Fig.algorithmpower} shows the performance of the proposed algorithm from the perspective of power minimization. In the early stage of training, the agent executes fairly random actions, which can be considered as a chaotic system and its transmit power is relatively high. By contrast, for well-trained scenarios, the joint DDPG-DQN scheme has achieved superior power minimization effects regardless of the action policy, which is in line with their trends in the reward analyses. The hybrid DDPG algorithm consumes more power than the DDPG-DQN algorithm, which is why it obtains less rewards. What is worth noting is that the baseline algorithm requires lower consumption than the hybrid DDPG algorithm. Recalling reward function \eqref{R} reveals that the solution provided by the DDPG agent does not satisfy \eqref{OPPD} even in some feasible conditions. Therefore, we can speculate that the DDPG algorithm relying on discretized actions can result in completely different actions on both sides of the threshold, which may lead to some action errors and may result in violations of the QoS constraint.

\begin{figure}[t!]
    \begin{center}
    \setlength{\belowcaptionskip}{-0.3cm}
        \includegraphics[width=0.45\textwidth]{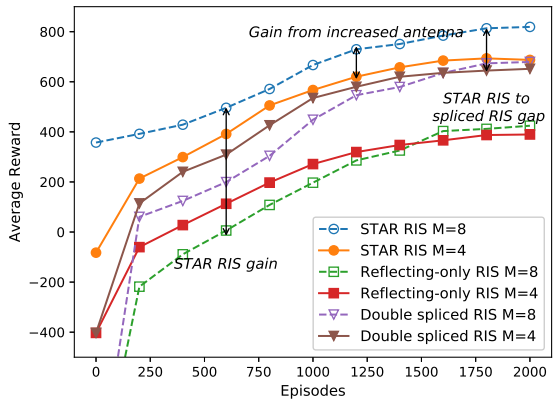}
        \caption{Performance comparison between STAR-RIS, reflecting-only RIS, and double spliced RIS}
        \label{Fig.breward}
    \end{center}
\end{figure}

The comparison between STAR-RIS, double spliced RIS, and reflecting-only RIS is presented in Fig. \ref{Fig.breward}. STAR-RIS has achieved the best overall performance in all training stages, and the double spliced RIS has about $7\%$ performance disadvantages over the STAR-RIS. This phenomenon confirms the conclusion that even if the transmitted and reflected signals of the STAR-RIS are constrained by each other, the STAR-RIS can provide further signal enhancement for users than double spliced RISs, since the STAR-RIS has higher multipath gain. The reflecting-only RIS has achieved a lower reward than the scheme having double-sides coverage, since the reflecting-only RIS is not capable of serving $\mathcal{T}$ users. In terms of the number of antennas, the simulation results prove that the proposed model and algorithm have general applicability. Furthermore, the multi-antenna gain of the STAR-RIS is higher than that of the reflecting-only RIS.

\begin{figure}[t!]
    \begin{center}
    \setlength{\belowcaptionskip}{-0.3cm}
        \includegraphics[width=0.45\textwidth]{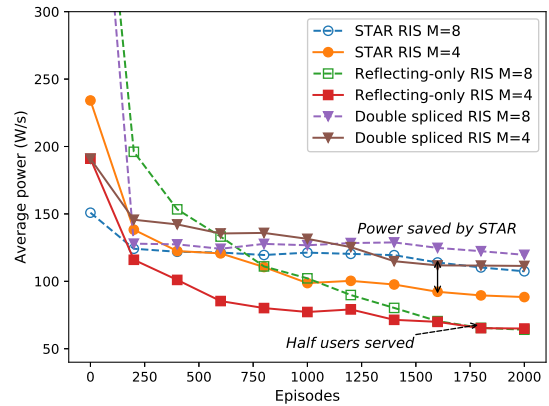}
        \caption{Power consumption of STAR-RIS, reflecting-only RIS, and double spliced RIS}
        \label{Fig.bpower}
    \end{center}
\end{figure}

The supporting evidence for Fig. \ref{Fig.breward} is provided by Fig. \ref{Fig.bpower}, which reveals the transmission power consumption of different types of RISs. The energy consumption characteristics shown by the STAR-RIS and double spliced RIS follow their reward trend. It can be observed that since the STAR-RIS can have a higher number of STAR elements to serve users, it obtained higher gains than the double spliced RIS. Thus, the BS dissipates less transmission power to meet the data rate requirements of users. It is worth paying attention to the fact that the reflecting-only RIS has the lowest power consumption in Fig. \ref{Fig.bpower}. However, this does not suggest that it has an overall favorable energy efficiency, since it only serves about half the users. Based on similar logic, in the case of $M=8$, the power consumption seen in Fig. \ref{Fig.bpower} is higher than for $M=4$, since the feasibility of the constraint \eqref{OPPD} has changed due to the increased number of antennas.

\begin{figure}[t!]
    \begin{center}
    \setlength{\belowcaptionskip}{-0.3cm}
        \includegraphics[width=0.45\textwidth]{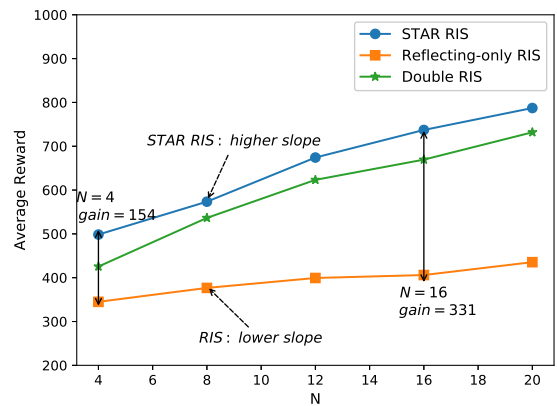}
        \caption{Performances against the number of STAR/reflecting elements}
        \label{Fig.MN}
    \end{center}
\end{figure}

Fig. \ref{Fig.MN} plots the reward against the number of STAR or reflection elements of (STAR) RISs. Upon increasing the number of elements, the rewards have also been improved to varying degrees, which is in line with the theoretical expectations of the diversity gain. The result indicates that the performance of the proposed algorithm is not significantly affected by the action dimension (element number). Finally, it can be observed in Fig. \ref{Fig.MN} that the gains obtained by the double-sided coverage of RISs are more significant than these of the single-sided RISs.
\section{Conclusions}\label{section:6}

A STAR-RIS assisted downlink network model was proposed and the effects of coupled transmission and reflection phase-shift model were considered for the STAR-RIS. Although the STAR-RIS expanded the service range of the reflecting-only RIS, optimizing the beamforming of the coupled transmission and reflection became a challenging problem, which required both continuous-valued and discrete-valued control. Thus, we developed a hybrid DDPG algorithm and a joint DDPG-DQN algorithm for jointly optimizing the active and passive beamforming to minimize the energy consumption. The analysis and simulation results indicated that 1) STAR-RIS exhibited superiority over the double spliced and the reflecting-only RISs in terms of consuming transmission energy; 2) The proposed hybrid DDPG algorithm and the DDPG-DQN algorithm have outperformed the conventional DDPG algorithm; 3) The DDPG-DQN algorithm achieved superior performance compared to the hybrid DDPG algorithm albeit at an increased complexity.


\renewcommand{\baselinestretch}{1}
\bibliography{STAR_RIS-REF}

\begin{thebibliography}{10}
\providecommand{\url}[1]{#1}
\csname url@samestyle\endcsname
\providecommand{\newblock}{\relax}
\providecommand{\bibinfo}[2]{#2}
\providecommand{\BIBentrySTDinterwordspacing}{\spaceskip=0pt\relax}
\providecommand{\BIBentryALTinterwordstretchfactor}{4}
\providecommand{\BIBentryALTinterwordspacing}{\spaceskip=\fontdimen2\font plus
\BIBentryALTinterwordstretchfactor\fontdimen3\font minus
  \fontdimen4\font\relax}
\providecommand{\BIBforeignlanguage}[2]{{%
\expandafter\ifx\csname l@#1\endcsname\relax
\typeout{** WARNING: IEEEtran.bst: No hyphenation pattern has been}%
\typeout{** loaded for the language `#1'. Using the pattern for}%
\typeout{** the default language instead.}%
\else
\language=\csname l@#1\endcsname
\fi
#2}}
\providecommand{\BIBdecl}{\relax}
\BIBdecl

\bibitem{8922634}
C.~Xu, N.~Ishikawa, R.~Rajashekar, S.~Sugiura, R.~G. Maunder, Z.~Wang, L.-L.
  Yang, and L.~Hanzo, ``Sixty years of coherent versus non-coherent tradeoffs
  and the road from 5{G} to wireless futures,'' \emph{IEEE Access}, vol.~7, pp.
  178\,246--178\,299, Dec. 2019.

\bibitem{9139273}
X.~{Mu}, Y.~{Liu}, L.~{Guo}, J.~{Lin}, and N.~{Al-Dhahir}, ``Exploiting
  intelligent reflecting surfaces in {NOMA} networks: Joint beamforming
  optimization,'' \emph{IEEE Trans. Wirel. Commun.s}, vol.~19, no.~10, pp.
  6884--6898, Oct. 2020.

\bibitem{9086766}
M.~A. {ElMossallamy}, H.~{Zhang}, L.~{Song}, K.~G. {Seddik}, Z.~{Han}, and
  G.~Y. {Li}, ``Reconfigurable intelligent surfaces for wireless
  communications: Principles, challenges, and opportunities,'' \emph{IEEE
  Trans. Cogn. Commun. Netw.}, vol.~6, no.~3, pp. 990--1002, Sept. 2020.

\bibitem{9521988}
W.~Zhang and W.~P. Tay, ``Cost-efficient {RIS}-aided channel estimation via
  rank-one matrix factorization,'' \emph{IEEE Wirel. Commun. Lett.}, vol.~10,
  no.~11, pp. 2562--2566, Nov. 2021.

\bibitem{9520407}
M.~Naderi~Soorki, W.~Saad, M.~Bennis, and C.~S. Hong, ``Ultra-reliable indoor
  millimeter wave communications using multiple artificial intelligence-powered
  intelligent surfaces,'' \emph{IEEE Trans. Commun.}, vol.~69, no.~11, pp.
  7444--7457, Nov. 2021.

\bibitem{8741198}
C.~Huang, A.~Zappone, G.~C. Alexandropoulos, M.~Debbah, and C.~Yuen,
  ``Reconfigurable intelligent surfaces for energy efficiency in wireless
  communication,'' \emph{IEEE Trans. Wirel. Commun.}, vol.~18, no.~8, pp.
  4157--4170, Aug. 2019.

\bibitem{alamzadeh2021reconfigurable}
I.~Alamzadeh, G.~C. Alexandropoulos, N.~Shlezinger, and M.~F. Imani, ``A
  reconfigurable intelligent surface with integrated sensing capability,''
  \emph{Sci. rep.}, vol.~11, no.~1, pp. 1--10, 2021.

\bibitem{9593172}
H.~Zhang, N.~Shlezinger, I.~Alamzadeh, G.~C. Alexandropoulos, M.~F. Imani, and
  Y.~C. Eldar, ``Channel estimation with simultaneous reflecting and sensing
  reconfigurable intelligent metasurfaces,'' in \emph{IEEE 22nd International
  Workshop SPAWC}, Sep. 2021, pp. 536--540.

\bibitem{9528041}
K.~Keykhosravi, M.~F. Keskin, S.~Dwivedi, G.~Seco-Granados, and H.~Wymeersch,
  ``Semi-passive {3D} positioning of multiple {RIS}-enabled users,'' \emph{IEEE
  Trans. Vehi. Technol.}, vol.~70, no.~10, pp. 11\,073--11\,077, Oct. 2021.

\bibitem{9530717}
S.~Basharat, S.~Ali~Hassan, H.~Pervaiz, A.~Mahmood, Z.~Ding, and M.~Gidlund,
  ``Reconfigurable intelligent surfaces: Potentials, applications, and
  challenges for 6{G} wireless networks,'' \emph{IEEE Wirel. Commun.}, pp.
  1--8, 2021.

\bibitem{9122596}
S.~Gong, X.~Lu, D.~T. Hoang, D.~Niyato, L.~Shu, D.~I. Kim, and Y.-C. Liang,
  ``Toward smart wireless communications via intelligent reflecting surfaces: A
  contemporary survey,'' \emph{IEEE Commun. Surv. Tutor.}, vol.~22, no.~4, pp.
  2283--2314, 2020.

\bibitem{mu2021simultaneously}
X.~Mu, Y.~Liu, L.~Guo, J.~Lin, and R.~Schober, ``Simultaneously transmitting
  and reflecting ({STAR}) {RIS} aided wireless communications,'' \emph{IEEE
  Trans. Wirel. Commun.}, vol. early access, Oct. 2021.

\bibitem{9201413}
S.~Zeng, H.~Zhang, B.~Di, Z.~Han, and L.~Song, ``Reconfigurable intelligent
  surface ({RIS}) assisted wireless coverage extension: {RIS} orientation and
  location optimization,'' \emph{IEEE Commun. Lett.}, vol.~25, no.~1, pp.
  269--273, Jan. 2021.

\bibitem{zhu2014dynamic}
B.~O. Zhu, K.~Chen, N.~Jia, L.~Sun, J.~Zhao, T.~Jiang, and Y.~Feng, ``Dynamic
  control of electromagnetic wave propagation with the equivalent principle
  inspired tunable metasurface,'' \emph{Sci. rep.}, vol.~4, no.~1, pp. 1--7,
  May. 2014.

\bibitem{7274678}
B.~O. Zhu and Y.~Feng, ``Passive metasurface for reflectionless and arbitary
  control of electromagnetic wave transmission,'' \emph{IEEE Trans. Antennas
  Propag.}, vol.~63, no.~12, pp. 5500--5511, Dec. 2015.

\bibitem{9437234}
J.~Xu, Y.~Liu, X.~Mu, and O.~A. Dobre, ``{STAR-RIS}s: Simultaneous transmitting
  and reflecting reconfigurable intelligent surfaces,'' \emph{IEEE Commun.
  Lett.}, vol.~25, no.~9, pp. 3134--3138, May. 2021.

\bibitem{9200683}
S.~Zhang, H.~Zhang, B.~Di, Y.~Tan, Z.~Han, and L.~Song, ``Beyond intelligent
  reflecting surfaces: Reflective-transmissive metasurface aided communications
  for full-dimensional coverage extension,'' \emph{IEEE Trans. Veh. Technol.},
  vol.~69, no.~11, pp. 13\,905--13\,909, Nov. 2020.

\bibitem{9462949}
C.~Wu, Y.~Liu, X.~Mu, X.~Gu, and O.~A. Dobre, ``Coverage characterization of
  {STAR-RIS} networks: {NOMA} and {OMA},'' \emph{IEEE Commun. Lett.}, vol.~25,
  no.~9, pp. 3036--3040, Sep. 2021.

\bibitem{liu2021simultaneously}
Y.~Liu, X.~Mu, R.~Schober, and H.~V. Poor, ``Simultaneously transmitting and
  reflecting ({STAR})-{RIS}s: A coupled phase-shift model,'' \emph{arXiv
  preprint arXiv:2110.02374}, 2021.

\bibitem{9115725}
S.~Abeywickrama, R.~Zhang, Q.~Wu, and C.~Yuen, ``Intelligent reflecting
  surface: Practical phase shift model and beamforming optimization,''
  \emph{IEEE Trans. Commun.}, vol.~68, no.~9, pp. 5849--5863, Sep. 2020.

\bibitem{wang2020thirty}
J.~Wang, C.~Jiang, H.~Zhang, Y.~Ren, K.-C. Chen, and L.~Hanzo, ``Thirty years
  of machine learning: The road to {P}areto-optimal wireless networks,''
  \emph{IEEE Commun. Surv. Tutor.}, vol.~22, no.~3, pp. 1472--1514, Jan. 2020.

\bibitem{9367208}
S.~Gao, P.~Dong, Z.~Pan, and G.~Y. Li, ``Deep multi-stage {CSI} acquisition for
  reconfigurable intelligent surface aided {MIMO} systems,'' \emph{IEEE Commun.
  Lett.}, vol.~25, no.~6, pp. 2024--2028, June 2021.

\bibitem{9274528}
H.~Song, M.~Zhang, J.~Gao, and C.~Zhong, ``Unsupervised learning-based joint
  active and passive beamforming design for reconfigurable intelligent surfaces
  aided wireless networks,'' \emph{IEEE Commun. Lett.}, vol.~25, no.~3, pp.
  892--896, Mar. 2021.

\bibitem{9110869}
C.~Huang, R.~Mo, and C.~Yuen, ``Reconfigurable intelligent surface assisted
  multiuser {MISO} systems exploiting deep reinforcement learning,'' \emph{IEEE
  J. Sel. Areas Commun.}, vol.~38, no.~8, pp. 1839--1850, Aug. 2020.

\bibitem{9206080}
H.~Yang, Z.~Xiong, J.~Zhao, D.~Niyato, L.~Xiao, and Q.~Wu, ``Deep reinforcement
  learning-based intelligent reflecting surface for secure wireless
  communications,'' \emph{IEEE Trans. Wirel. Commun.}, vol.~20, no.~1, pp.
  375--388, Jan. 2021.

\bibitem{9174801}
X.~Liu, Y.~Liu, Y.~Chen, and H.~V. Poor, ``{RIS} enhanced massive
  non-orthogonal multiple access networks: Deployment and passive beamforming
  design,'' \emph{IEEE J. Sel. Areas Commun.}, vol.~39, no.~4, pp. 1057--1071,
  Apr. 2021.

\bibitem{9371415}
M.~Samir, M.~Elhattab, C.~Assi, S.~Sharafeddine, and A.~Ghrayeb, ``Optimizing
  age of information through aerial reconfigurable intelligent surfaces: A deep
  reinforcement learning approach,'' \emph{IEEE Trans. Veh. Tech.}, vol.~70,
  no.~4, pp. 3978--3983, Mar. 2021.

\bibitem{ni2021star}
W.~Ni, Y.~Liu, Y.~C. Eldar, Z.~Yang, and H.~Tian, ``{STAR-RIS} enabled
  heterogeneous networks: Ubiquitous {NOMA} communication and pervasive
  federated learning,'' \emph{arXiv preprint arXiv:2106.08592}, 2021.

\bibitem{neunert2020continuous}
M.~Neunert, A.~Abdolmaleki, M.~Wulfmeier, T.~Lampe, T.~Springenberg, R.~Hafner,
  F.~Romano, J.~Buchli, N.~Heess, and M.~Riedmiller, ``Continuous-discrete
  reinforcement learning for hybrid control in robotics,'' in \emph{Proceedings
  of the CoRL}, vol. 100, 30 Oct--01 Nov 2020, pp. 735--751.

\bibitem{delalleau2019discrete}
O.~Delalleau, M.~Peter, E.~Alonso, and A.~Logut, ``Discrete and continuous
  action representation for practical {RL} in video games,'' \emph{arXiv
  preprint arXiv:1912.11077}, 2019.

\bibitem{li2021hyar}
B.~Li, H.~Tang, Y.~Zheng, J.~Hao, P.~Li, Z.~Wang, Z.~Meng, and L.~Wang, ``Hyar:
  Addressing discrete-continuous action reinforcement learning via hybrid
  action representation,'' \emph{arXiv preprint arXiv:2109.05490}, 2021.

\bibitem{9110912}
S.~Zhang and R.~Zhang, ``Capacity characterization for intelligent reflecting
  surface aided {MIMO} communication,'' \emph{IEEE J. Sel. Areas Commun.},
  vol.~38, no.~8, pp. 1823--1838, Aug. 2020.

\bibitem{TR36873}
``Study on 3d channel model for lte (3gpp tr 36.873 release 12),'' \emph{3rd
  Generation Partnership Project}, Jan. 2018.

\bibitem{lillicrap2015continuous}
T.~P. Lillicrap, J.~J. Hunt, A.~Pritzel, N.~Heess, T.~Erez, Y.~Tassa,
  D.~Silver, and D.~Wierstra, ``Continuous control with deep reinforcement
  learning,'' \emph{arXiv preprint arXiv:1509.02971}, 2015.

\bibitem{li2021joint}
Y.~Li, A.~H. Aghvami, and Y.~Deng, ``Joint resource block and beamforming
  optimization for cellular-connected {UAV} networks: A hybrid {D3QN-DDPG}
  approach,'' \emph{arXiv preprint arXiv:2102.13222}, 2021.

\bibitem{9372298}
A.~Feriani and E.~Hossain, ``Single and multi-agent deep reinforcement learning
  for {AI}-enabled wireless networks: A tutorial,'' \emph{IEEE Commun. Surv.
  Tutor.}, vol.~23, no.~2, pp. 1226--1252, Mar. 2021.

\bibitem{9312988}
I.~Abraham, A.~Prabhakar, and T.~D. Murphey, ``An ergodic measure for active
  learning from equilibrium,'' \emph{IEEE Trans. Autom. Sci. Eng.}, vol.~18,
  no.~3, pp. 917--931, July 2021.

\bibitem{8103164}
K.~Arulkumaran, M.~P. Deisenroth, M.~Brundage, and A.~A. Bharath, ``Deep
  reinforcement learning: A brief survey,'' \emph{IEEE Signal Process. Mag.},
  vol.~34, no.~6, pp. 26--38, Nov. 2017.

\bibitem{qiu2019deep}
C.~Qiu, Y.~Hu, Y.~Chen, and B.~Zeng, ``Deep deterministic policy gradient
  ({DDPG})-based energy harvesting wireless communications,'' \emph{IEEE
  Internet Things J.}, vol.~6, no.~5, pp. 8577--8588, Oct. 2019.

\bibitem{9354068}
C.~Li, J.~Xia, F.~Liu, D.~Li, L.~Fan, G.~K. Karagiannidis, and A.~Nallanathan,
  ``Dynamic offloading for multiuser muti-{CAP MEC} networks: A deep
  reinforcement learning approach,'' \emph{IEEE Trans. Vehi. Technol.},
  vol.~70, no.~3, pp. 2922--2927, Mar. 2021.

\end{thebibliography}
\bibliographystyle{IEEEtran}

\end{document}